\documentclass[12pt]{article}

\usepackage{graphicx,color}
\def\beq{\begin{equation}}
\def\eeq#1{\label{#1}\end{equation}}
\def\eeqn{\end{equation}}

\def\beqa{\begin{eqnarray}}
\def\eeqa#1{\label{#1}\end{eqnarray}}
\def\eeqan{\end{eqnarray}}

\let\bar=\overbar

\def\eg{{\it e.g.}}

\def\Dslash{\not{\hbox{\kern-4pt $D$}}}
\def\dslash{\not{\hbox{\kern-2pt $\del$}}}

\def\msb{{\bar{\ssstyle M \kern -1pt S}}}

\def\pbinv{pb$^{-1}$}
\def\fbinv{fb$^{-1}$}

\def\calB{{\cal B}}

\def\eg{{\it e.g.}}

\def\Ecm{E_{cm}}

\def\elp{e^+}
\def\elm{e^-}

\def\pip{\pi^+}
\def\pim{\pi^-}
\def\piz{\pi^0}
\def\pipm{\pi^\pm}

\def\Kp{K^+}
\def\Km{K^-}
\def\Kz{K^0}
\def\Kpstar{K^{*+}}
\def\Kmstar{K^{*-}}
\def\Kzbar{\bar{K}^0}

\def\KS{K^0_S}
\def\KL{K^0_L}
\def\Kpm{K^\pm}
\def\Kmp{K^\mp}
\def\Kpmstar{K^{*\pm}}

\def\Dp{D^+}
\def\Dm{D^-}
\def\Dz{D^0}
\def\Dzbar{\bar{D}^0}
\def\Dstar{D^*}
\def\Dstarbar{\bar{D}^*}

\def\Dhat{\hat{D}}
\def\Dzhat{\Dhat^0}
\def\Dzhatstar{\Dhat^{0(*)}}

\def\Ds{D_s}
\def\Dsp{D_s^+}
\def\Dsm{D_s^-}
\def\Dsstar{D_s^*}

\def\Dspm{D_s^\pm}

\def\Dsmpbar{\bar{D}_s^\mp}
\def\Dspmstar{D_s^{*\pm}}
\def\Dsmpstar{D_s^{*\mp}}

\def\Bm{B^-}
\def\Bp{B^+}
\def\Bpm{B^\pm}

\def\Dbar{\bar{D}}

\def\NDDbar{N_{D\Dbar}}

\def\jbar{\bar{\jmath}}

\def\Bi{\calB_i}

\def\Ni{N_i}
\def\effi{\epsilon_i}
\def\Bjbar{\calB_{\jbar}}
\def\Njbar{N_{\jbar}}
\def\effjbar{\epsilon_{\jbar}}
\def\Nijbar{N_{i\jbar}}
\def\effijbar{\epsilon_{i\jbar}}

\def\Mbc{M_{BC}}

\def\MDs{M(\Ds)}

\def\psidprime{\psi(3770)}

\def\Dzkpi{D^0\to K^-\pi^+}
\def\Dzkpipiz{D^0\to K^-\pi^+\pi^0}
\def\Dzkpipipi{D^0\to K^-\pi^+\pi^+\pi^-}
\def\Dpkpipi{D^+\to K^-\pi^+\pi^+}
\def\Dpkpipipiz{D^+\to K^-\pi^+\pi^+\pi^0}
\def\Dpkspi{D^+\to K^0_S\pi^+}
\def\Dpkspipiz{D^+\to K^0_S\pi^+\pi^0}
\def\Dpkspipipi{D^+\to K^0_S\pi^+\pi^+\pi^-}
\def\Dpkkpi{D^+\to K^+K^-\pi^+}

\def\sbar{\bar{s}}

\def\Vcs{V_{cs}}
\def\Vcd{V_{cd}}

\def\ED{E(D)}
\def\Ez{E_0}

\def\phithree{\phi_3}

\catcode`\@=11 % @ signs can be used
\newcommand{\Begitem}{\begin{list}
{\csname\@itemitem\endcsname}
{\ifnum \@itemdepth >3 \@toodeep\else \advance\@itemdepth 1 
\edef\@itemitem{labelitem\romannumeral\the\@itemdepth} 
  \parsep  2pt plus 1pt minus 1pt 
  \parskip 0pt plus 1pt minus 1pt 
  \topsep  0pt plus 1pt minus 1pt             
  \itemsep 0pt plus 1pt minus 1pt \fi}}
\newcommand{\Enditem}{\end{list}}
\catcode`\@=12 % @ signs cannot be used

\newcommand{\Arraystretch}{1.3}
\newcommand{\Begtabular}[1]{\renewcommand{\arraystretch}{\Arraystretch}
                         \begin{center}\begin{tabular}{#1}}
\newcommand{\Endtabular}{\hline    \end{tabular}\end{center}}

\def\ltsim{~\raisebox{-.55ex}{\rlap{$\sim$}} \raisebox{.45ex}{$<$}~}

\def\Fig#1{Figure~\ref{#1}}
\def\Tab#1{Table~\ref{#1}}
\def\Sec#1{Section~\ref{#1}}

\def\apscredit#1#2{Reprinted figure with permission from Ref.~\cite{#1}.  Copyright #2 by the American Physical Society.}

\def\plbcredit#1#2{Reprinted from Ref.~\cite{#1}, Copyright #2, with permission from Elsevier}

\def\Title#1{\begin{center} {\Large {\bf #1} } \end{center}}

\begin{document}

\Title{\boldmath Hadronic $D$ Decays and Dalitz Analyses}

\begin{center}{\large \bf Contribution to the proceedings of HQL06,\\
Munich, October 16th-20th 2006}\end{center}

\bigskip\bigskip

%+\addtocontents{toc}{{\it D. Reggiano}}
%+\label{ReggianoStart}

\begin{raggedright}  

{\it David G. Cassel\index{Cassel, D.G.}\\
Laboratory for Elementary-Particle Physics\\
Cornell University\\
Ithaca, NY 14853 USA}
\bigskip\bigskip
\end{raggedright}

\section{Introduction}

Recently several factors converged to ignite a renaissance in charm physics.  These factors include: 
\Begitem
\item the need for precision measurements from the charm sector to interpret  quantitatively $CP$ violation results from the beauty sector, 
\item technical and computational developments in Lattice QCD that led to precise calculations that can face precision experimental challenges in the charm sector,
\item  discovery of new -- generally unanticipated -- charm meson states (\eg, $D^*_{s0}(2317)$ and $D_{s1}(2460)$), and
\item substantial increases in the precision and reach of charm decay measurements due to production of much larger charm data sets (particularly at CESR,  KEKB, and PEP~II) and excellent detectors in the experiments at these facilities.
\Enditem 
  In this report I describe recent results in hadronic decays of the stable $D$ mesons ($\Dz$, $\Dp$, and $\Ds$) and Dalitz analyses of resonance structure in their hadronic decays.  These results come from experiments at electron-positron colliders (BaBar, Belle, and CLEO-c) and fixed target experiments (E791 and FOCUS).   

\section{\boldmath Absolute $\Dz$ and $\Dp$ Branching Fractions\label{sec:dabsolute}}

CLEO is providing new precise measurements of absolute $\Dz$ and $\Dp$ branching fractions using $\Dp$ and $\Dz$ decays from $\elp\elm \to \psidprime \to \Dp \Dm$~or~$\Dz\Dzbar$.  The mass of the $\psidprime$ is below the threshold for $D\Dbar\pi$ decays, so no additional pions are produced.  Furthermore, the multiplicities of $\Dz$ and $\Dp$ decays are low, so events are extremely clean. 
CLEO measures leptonic, semileptonic, and key hadronic branching fractions using a double tagging technique pioneered by MARK~III \cite{markiii-1,markiii-2}.  Most other $D$ branching fractions \cite{pdg:06} are measured relative to a reference mode, usually $\Dz\to\Km\pip$ or $\Dp\to\Km\pip\pip$.  CLEO has  published absolute branching fractions for key Cabibbo-Favored Decays (CFD) to hadrons obtained from 56~\pbinv of data \cite{cleo:dhad}.  I report here a preliminary update from 281~\pbinv\ of data; this is the first public presentation of these results.  Some other branching ratios utilizing the 281~\pbinv\ data sample have been  published or submitted for publication.

The MARK~III double tag technique derives absolute branching fractions from measurements of Single Tag (ST) and Double Tag (DT) yields.  In ST events, the $D(\Dbar)$ is reconstructed in a specific final state, while the decay products of the $\Dbar(D)$ are not observed, \eg, $D \to i$ and $\Dbar \to \bar{X}$.  In DT events, both the $D$ and the $\Dbar$ are reconstructed in specific final states, \eg, $D \to i$ and $\Dbar \to \jbar$.  
ST and DT yields are given by $\Ni = \NDDbar\; \Bi\,\effi$,  $\Njbar = \NDDbar\; \Bjbar\,\effjbar$, and 
$\Nijbar = \NDDbar\; \Bi\,\Bjbar\,\effijbar$, respectively, where: $i$ and $\jbar$ are the decay modes of the $D$ and $\Dbar$, respectively; $\Bi$, and $\Bjbar$ are the corresponding branching fractions; $\effi$, $\effjbar$, and $\effijbar$ are the ST and DT efficiencies for these modes;  $\Ni$, $\Njbar$, and $\Nijbar$ are the corresponding ST and DT yields; and $\NDDbar$ is the number of $\Dz\Dzbar$ or $\Dp\Dm$ events produced in the experiment.
Branching fractions and $\NDDbar$ can then be obtained from 
\[ \Bi = {\Nijbar \over \Njbar}~{\effjbar \over \effijbar}\quad
\textrm{and}\quad  \NDDbar = {\Ni\Njbar \over \Nijbar}\; {\effijbar \over \effi\effjbar} \]
where $\effjbar$ is the efficiency for detecting $\Dbar\to\jbar$ ST events, and $CP$ symmetry has been assumed so $\calB_{\bar{\imath}} = \Bi$.  Obviously the luminosity is not required to determine $\Bi$ or $\NDDbar$, but the cross section for $\elp\elm \to \psidprime \to D\Dbar$ can be determined from $\NDDbar$ and the luminosity.  Furthermore, $\effijbar \approx \effi\effjbar$ so the branching fractions and number of events obtained by this method are relatively insensitive to uncertainties in the efficiencies. 
 
CLEO utilizes a $\chi^2$ fit~\cite{wsun} that obtains all $\Dz$ and $\Dp$ branching fractions, as well as the numbers of $\Dz\Dzbar$ and $\Dp\Dm$ events from a simultaneous fit to all ST and DT $\Dz$ and $\Dp$ yields.  All statistical and systematic uncertainties and their correlations are properly taken into account in this fit.

\begin{figure}[htb]
\begin{center}
\includegraphics[width=5.0in]{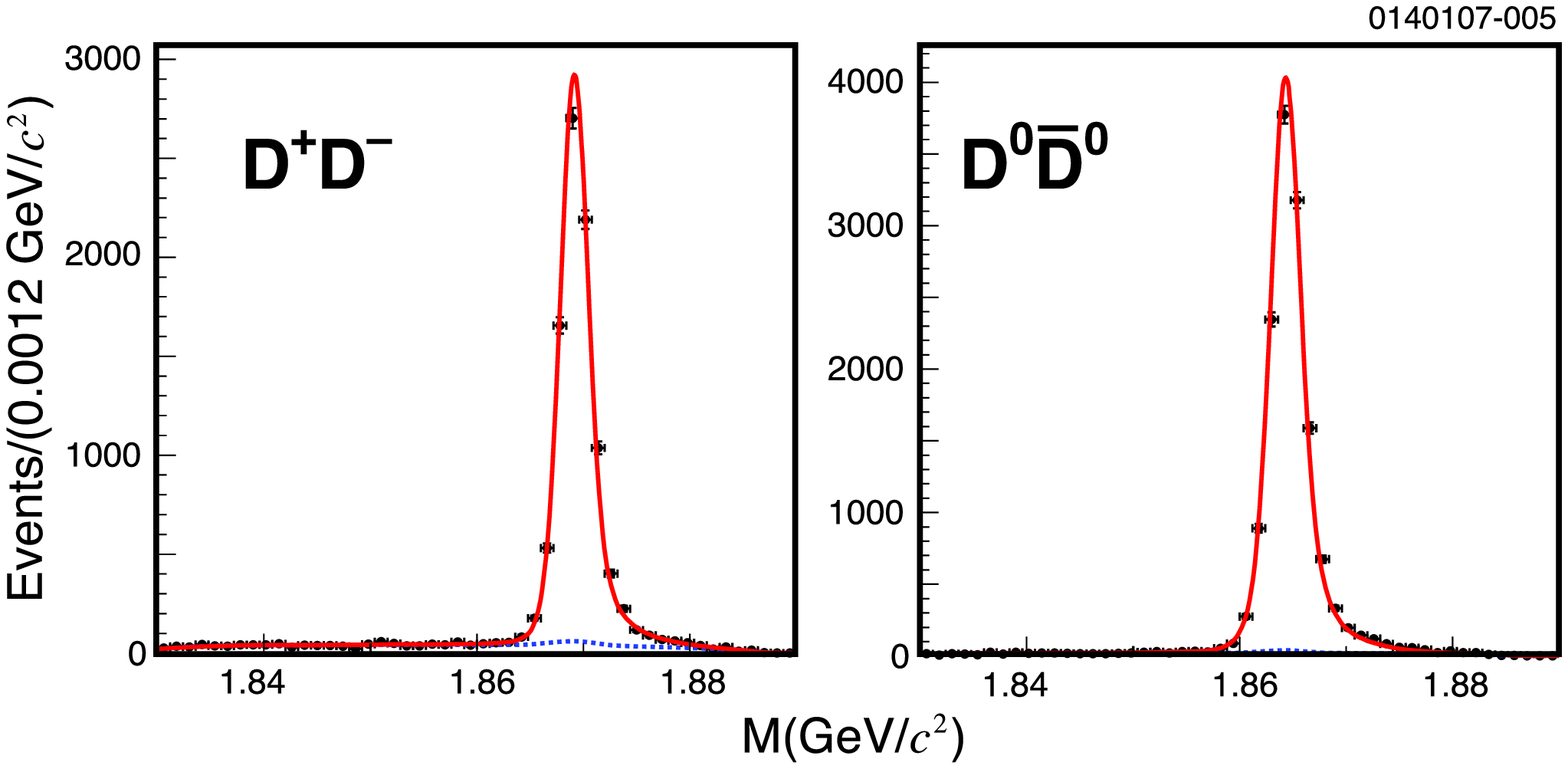}
\caption{Histograms of $\Mbc$ distributions of $D$ and $\Dbar$ candidates in double-tag events from 281~\pbinv\ of CLEO-c $\psidprime$ data; (left) $\Dp\Dm$ DT candidates and (right) $\Dz\Dzbar$ candidates.}
\label{fig:dthists}
\end{center}
\end{figure}

Candidate $D$ or $\Dbar$ mesons are selected using mode-dependent requirements on $\Delta E \equiv \ED - \Ez$, the difference between the measured energies $\ED$ of the candidate and the beam energy $\Ez$.   
Then ST and DT yields are obtained from fits to beam constrained mass $\Mbc$ distributions, where the beam energy $\Ez$ is substituted for the measured energy $\ED$ of a $D$ or $\Dbar$ candidate.  Substitution of $\Ez$ for $\ED$ improves substantially the mass resolution of the $D$ candidates.  The $\Mbc$ distributions for the DT events are illustrated in \Fig{fig:dthists}. Clearly these $\Mbc$ distributions are very clean with little background.  Summed over all modes, the yields obtained from the CLEO-c 281~\pbinv\ sample are 230,225 ST and $13,575 \pm 120$ DT $\Dz\Dzbar$ events, and 167,086 ST and $8,867 \pm 97$ DT $\Dp\Dm$ events.  With these yields, systematic errors dominate; to be conservative for these preliminary results, most systematic errors are essentially the same as those determined for the 56~\pbinv\ results.  Intensive study of the systematic uncertainties are underway and CLEO expects to be able to substantially  improve most of them.  

\begin{table}[htb]
\vspace*{-0.3in}
\begin{center}
\Begtabular{l|cc}
Mode & $\calB$ (\%) \\
\hline
%$\NDzDzbar$               & $(2.01\pm 0.04\pm 0.02)\times 10^5$  \\
$\Dzkpi$        & $3.87\pm 0.04\pm 0.08$         \\
$\Dzkpipiz$     & $14.6\pm 0.1\pm 0.4$            \\
$\Dzkpipipi$    & $8.3\pm 0.1\pm 0.3$           \\
\hline
%$\NDpDm$                  & $(1.56\pm 0.04\pm 0.01)\times 10^5$ \\
$\Dpkpipi$      & $9.2\pm 0.1\pm 0.2$            \\
$\Dpkpipipiz$   & $6.0\pm 0.1\pm 0.2$            \\
$\Dpkspi$       & $1.55\pm 0.02\pm 0.05$         \\
$\Dpkspipiz$    & $7.2\pm 0.1\pm 0.3$            \\
$\Dpkspipipi$   & $3.13\pm 0.05\pm 0.14$            \\
$\Dpkkpi$       & $0.93\pm 0.02\pm 0.03$         \\
\Endtabular
\caption{Preliminary CLEO results for hadronic $\Dz$ and $\Dp$ branching fractions, with their statistical and systematic errors.  These results come from 281~\pbinv\ of CLEO-c $\psidprime$ data, with conservative systematic errors.}
\label{tab:BDKnpi}
\end{center}
\end{table}

\begin{figure}[htb]
\begin{center}
\includegraphics[width=2.9in]{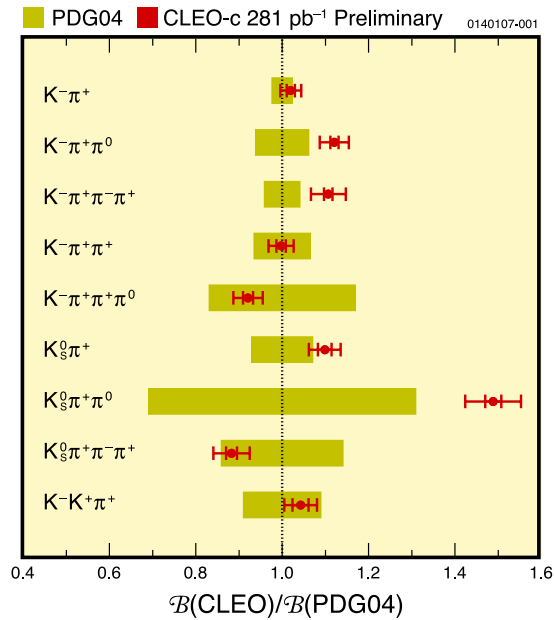}
\caption{Ratio $\calB(\textrm{CLEO})/\calB(\textrm{PDG04})$
of preliminary CLEO-c hadronic $\Dz$ and $\Dp$ branching fractions to the 2004 PDG averages.  The widths of the PDG bars correspond to the errors in those averages.  The CLEO-c points have statistical and systematic error bars.}
\label{fig:BDKnpi-PDG}
\end{center}
\end{figure}

The preliminary CLEO results for three $\Dz$ and six $\Dp$ hadronic decay modes are given in \Tab{tab:BDKnpi} and compared to PDG04~\cite{pdg:04} averages in \Fig{fig:BDKnpi-PDG}.  It is clear that the preliminary CLEO results are substantial improvements over previous measurements.  PDG04 averages were used for the comparison because CLEO-c 56~\pbinv\ results were included in the PDG06 averages.  Final State Radiation (FSR) is included in the CLEO-c Monte Carlo (MC) simulations for determining  efficiencies.  If FSR had not been included in the simulations the branching fractions would decrease by $\ltsim 2$\%.  Most other measurements of hadronic $D$ branching fractions did not take FSR into account.

%\clearpage

\section{\boldmath Cabibbo Suppressed $D^0$ and $D^+$ Decays}

\subsection {\boldmath Singly-Cabibbo-Suppressed $\Dz$ and $\Dp$ Decays to Pions\label{sec:Dnpi}}

\begin{figure}[htb]
\begin{center}
\begin{minipage}[t]{2.9in}
\includegraphics[width=2.8in]{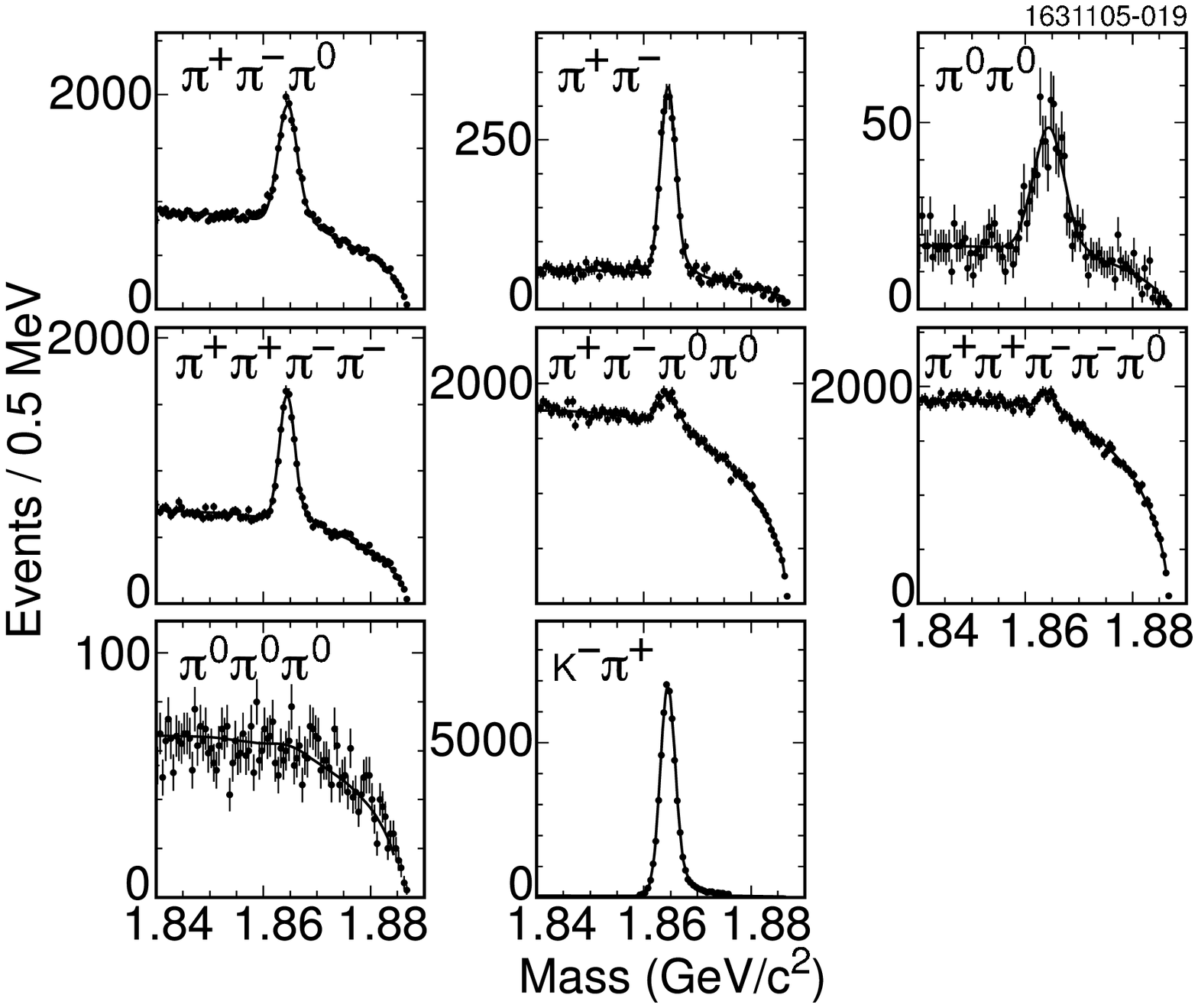}\hfill
\end{minipage}\hfill
\begin{minipage}[t]{2.9in}
\vspace*{-2.35in}
\includegraphics[width=2.8in]{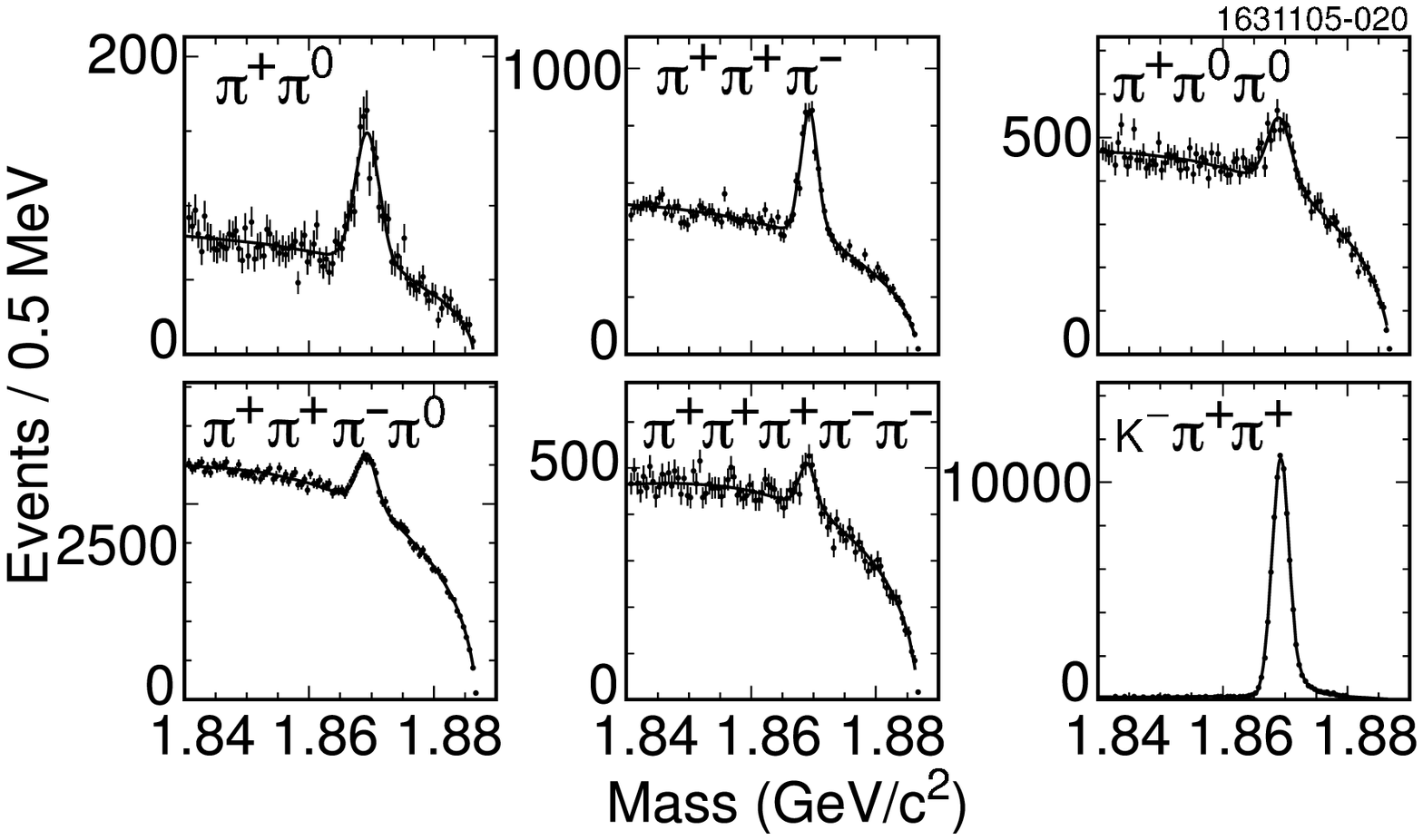}
\end{minipage}
\caption{Histograms of $\Mbc$ for (plots on the left) of $\Dz\to (n)\pi$ events and (plots on the right) $\Dp\to (n)\pi$ from the 281~\pbinv\ $\psidprime$ CLEO-c data sample.  Histograms of $\Mbc$ for the reference modes $\Dz\to\Km\pip$ and $\Dp\to\Km\pip\pip$ are included with the $D \to n(\pi)$ histograms on the left and right, respectively.}
\label{fig:Dnpi}
\end{center}
\end{figure}

\begin{table}[htb]
\begin{center}
\Begtabular{l|cc}
Mode & CLEO-c $\calB$ ($10^{-3}$) &  PDG04 $\calB$ ($10^{-3}$) \\ \hline
$\pip\pim$ & $1.39 \pm 0.04 \pm 0.03$ & $1.38 \pm 0.05$ \\
$\piz\piz$ & $0.79 \pm 0.05 \pm 0.04$ & $0.84 \pm 0.22$ \\
$\pip\pim\piz$ & $13.2 \pm 0.2 \pm 0.5$ & $11 \pm 4$ \\
$\pip\pim\pip\pim$ & $7.3 \pm 0.1 \pm 0.3$ & $7.3 \pm 0.5$ \\
$\pip\pim\piz\piz$ & $9.9 \pm 0.6 \pm 0.7$ & ---  \\
$\pip\pim\pip\pim\piz$ & $4.1 \pm 0.5  \pm 0.2$ & --- \\ \hline
$\pip\piz$ & $1.25 \pm 0.06 \pm 0.08$ & $1.33 \pm 0.22$ \\
$\pip\pip\pim$ & $3.35 \pm 0.10 \pm 0.20$ & $3.1 \pm 0.4$ \\
$\pip\piz\piz$ & $4.8 \pm 0.3 \pm 0.4$ & --- \\
$\pip\pip\pim\piz$ & $11.6 \pm 0.4 \pm 0.7$ & --- \\
$\pip\pim\pip\pim\pip$ & $1.60 \pm 0.18 \pm 0.17$ & $1.82 \pm 0.25$ \\
\Endtabular
\caption{CLEO branching fractions for $\Dz$ and $\Dp$ decays to multiple pions.}  \label{tab:Dnpi}
\end{center}
\end{table}

\begin{figure}[htb]
\begin{center}
\includegraphics[width=2.9in]{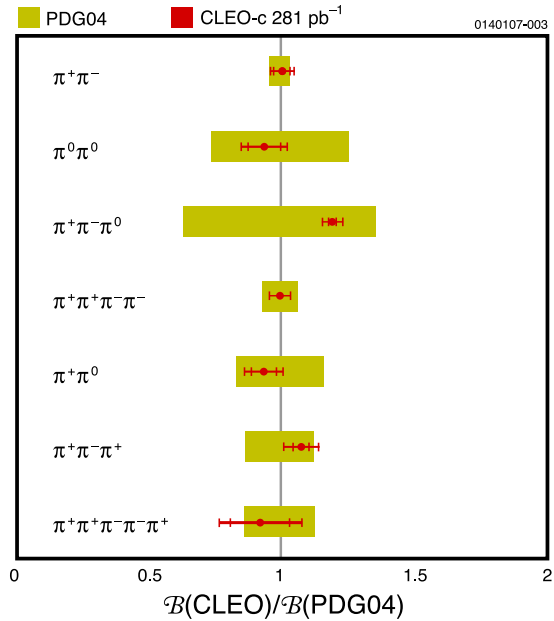}
\caption{Ratio $\calB(\textrm{CLEO})/\calB(\textrm{PDG06})$ of CLEO-c measurements of $D\to n(\pi)$ branching fractions to the 2006 PDG averages.  The widths of the PDG bars correspond to the errors in those averages. }
\label{fig:Dnpi-PDG}
\end{center}
\end{figure}

Using the full 281~\pbinv\ $\psidprime$ data sample, CLEO measured Singly-Cabibbo-Suppressed Decays (SCSD) of $\Dz$ and $\Dp$ to multipion final states \cite{cleo:Dnpi}.  The $\Mbc$ distributions are illustrated in \Fig{fig:Dnpi}.  
The branching fractions in \Tab{tab:Dnpi} were obtained from measurements of $\calB(\Dz\to n(\pi))/\calB(\Dz\to\Km\pip)$ and $\calB(\Dp\to n(\pi))/\calB(\Dp\to\Km\pip\pip)$.  The reference branching fractions used to determine the reported multipion branching fractions were $\calB(\Dz \to \Km\pip) = (3.84 \pm 0.07)$\% and $\calB(\Dp\to\Km\pip\pip) = (9.4 \pm 0.3)$\%.  These were obtained by averaging the 56~\pbinv\ CLEO-c results and the PDG04 averages.  These new CLEO measurements are compared to the PDG04~\cite{pdg:04} averages in \Fig{fig:Dnpi-PDG}.

As described in \Sec{sec:DpKppim}, BaBar has also reported a measurement of $\calB(\Dp\to\pip\piz)$ along with the first observation of the Doubly-Cabibbo-Suppressed-Decay\break (DCSD) branching fraction $\calB(\Dp\to\Kp\piz)$. The BaBar result for the ratio of $\calB(\Dp\to\pip\piz)$ to the reference branching fraction $\calB(\Dp\to\Km\pip\pip)$ is in excellent agreement with the CLEO measurement.  

The ratio of the isospin amplitudes $A_0$ and $A_2$ for $D\to\pi\pi$ decay and their phase difference can be determined from the values of $\calB(\pip\pim)$, $\calB(\piz\piz)$, and $\calB(\pip\piz)$ and the $\Dz$ and $\Dp$ decay widths~\cite{cleo:D2pi}.  Only $I = 0$ and 2 states are allowed for a two pion system, since an isovector state is forbidden.  Using $D$ lifetimes from PDG04~\cite{pdg:04} and the branching fractions in \Tab{tab:Dnpi}, CLEO finds $A_2/A_0 = 0.420 \pm 0.014 \pm 0.001$ and the relative phase $\delta = (86.4 \pm 2.8 \pm 3.3)^\circ$.  In contrast to the isospin amplitudes in $K\to\pi\pi$ decays, the two isospin amplitudes in $D\to\pi\pi$ decay are comparable~\cite{cleo:D2pi}.  Furthermore, the large phase difference between the two amplitudes indicates that final state interactions are important.

\subsubsection{Searches for $\eta$ and $\omega$ in multipion $\Dz$ and $\Dp$ decays}

\begin{figure}[htb]
\begin{center}
\includegraphics[width=3.0in]{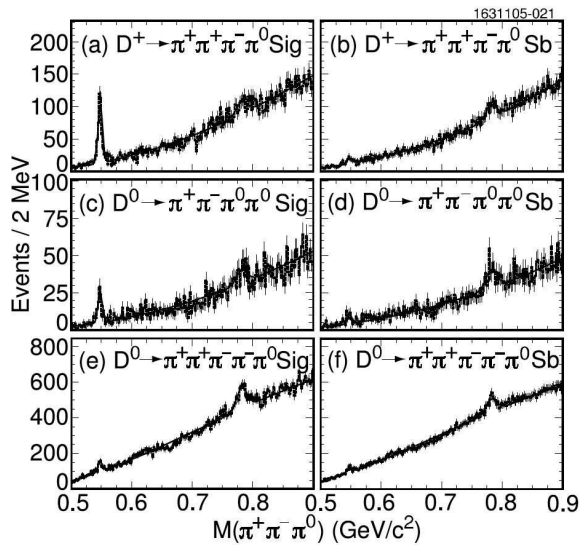}
\caption{$M(\pip\pim\piz)$ distributions for multipion $\Dp$ and $\Dz$ decays from CLEO-c data.}
\label{fig:Mpippimpiz}
\end{center}
\end{figure}

Untangling the resonant substructure of these decays with more than two pions would require systematic Dalitz analyses.  However, searches for $\eta$s and $\omega$s among the decay pions is a first step in this direction.  The technique used is illustrated in \Fig{fig:Mpippimpiz}, where  three-pion invariant masses, $M(\pip\pim\piz)$ are plotted for $D$ or $\Dbar$ candidates within the original $\Delta E$ requirement and in sidebands.  Although combinatorial backgrounds are large, there are significant signals above background peaks in: $\eta\piz$ in $\Dz\to\pip\pim\piz\piz$, $\eta\pip$ in  $\Dp\to\pip\pip\pim\piz$, and $\omega\pip\pim$ in $\Dz\to\pip\pim\pip\pim\pip$.  The branching fractions for these three channels and 90\% UL confidence intervals for the other three modes are given in \Tab{tab:Betaomega}.  The branching fractions for $\eta\pip$ and $\omega\pip\pim$ are significant fractions of the total branching fractions for the parent decays.

\begin{table}[htb]
\begin{center}
\Begtabular{l|cc}
Mode & $\calB$ ($10^{-3}$) &  PDG04 $\calB$ ($10^{-3}$) \\ \hline
$\eta\piz$ & $0.62 \pm 0.14 \pm 0.05$ & --- \\
$\eta\pip$ & $ 3.61 \pm 0.25 \pm 0.26$ & $3.0 \pm 0.6$ \\
$\eta\pip\pim$ & $ < 1.9$ (90\% CL) & --- \\ \hline
$\omega\piz$ & $ < 0.26$ (90\% CL) & --- \\
$\omega\pip$ & $ < 0.34$ (90\% CL) & --- \\
$\omega\pip\pim$ & $1.7 \pm 0.5 \pm 0.2$ & --- \\
\Endtabular
\caption{Branching fractions and upper limits for $\eta$ and $\omega$ production in multipion $\Dz$ and $\Dp$ decays from CLEO-c data.}
\label{tab:Betaomega}
\end{center}
\end{table}

\subsection{\boldmath Singly-Cabibbo-Suppressed Decays from BaBar and Belle}

\begin{figure}[htb]
\begin{center}
\includegraphics[width=2.7in]{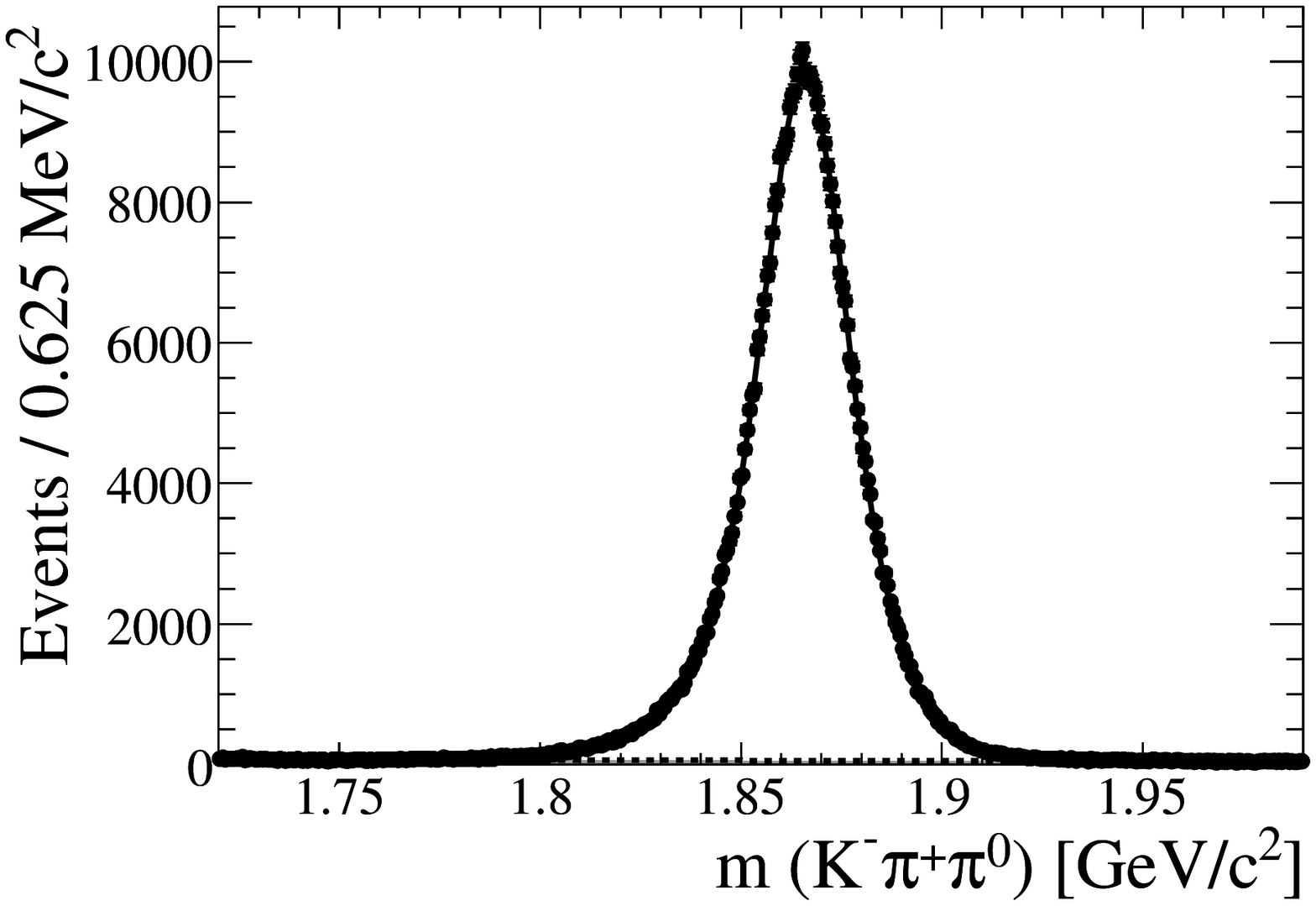}\hspace*{0.3in}
\includegraphics[width=2.7in]{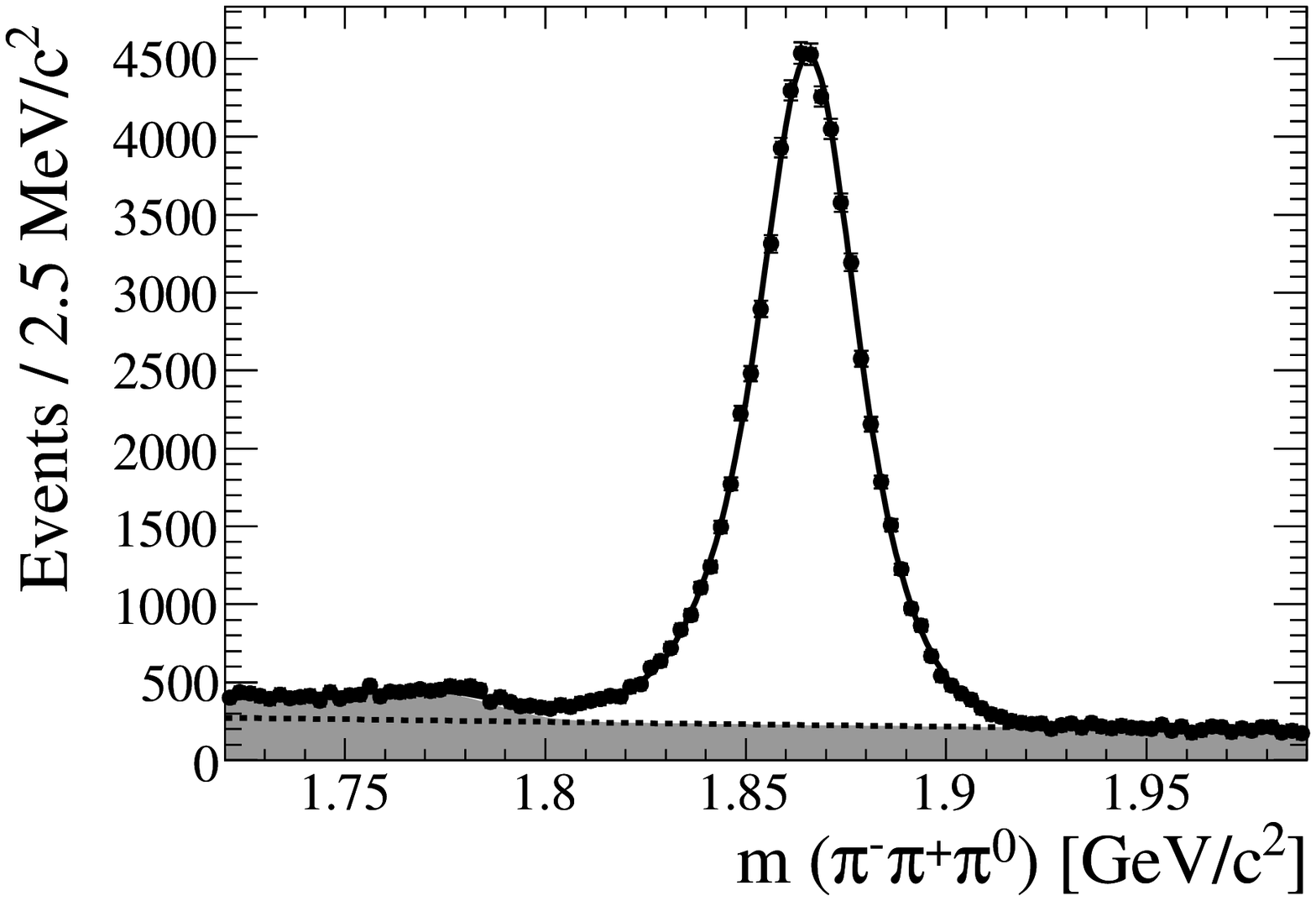}
\caption{BaBar data for (left) $\Dz\to\Km\pip\piz$ decays and (right) $\Dz\to\pip\pim\piz$ decays.~\cite{aps:babar:scsd,aps:extra}}
\label{fig:BaBar-Dzpipipi}
\end{center}
\end{figure}

\begin{figure}[htb]
\begin{center}
\includegraphics[height=2.7in]{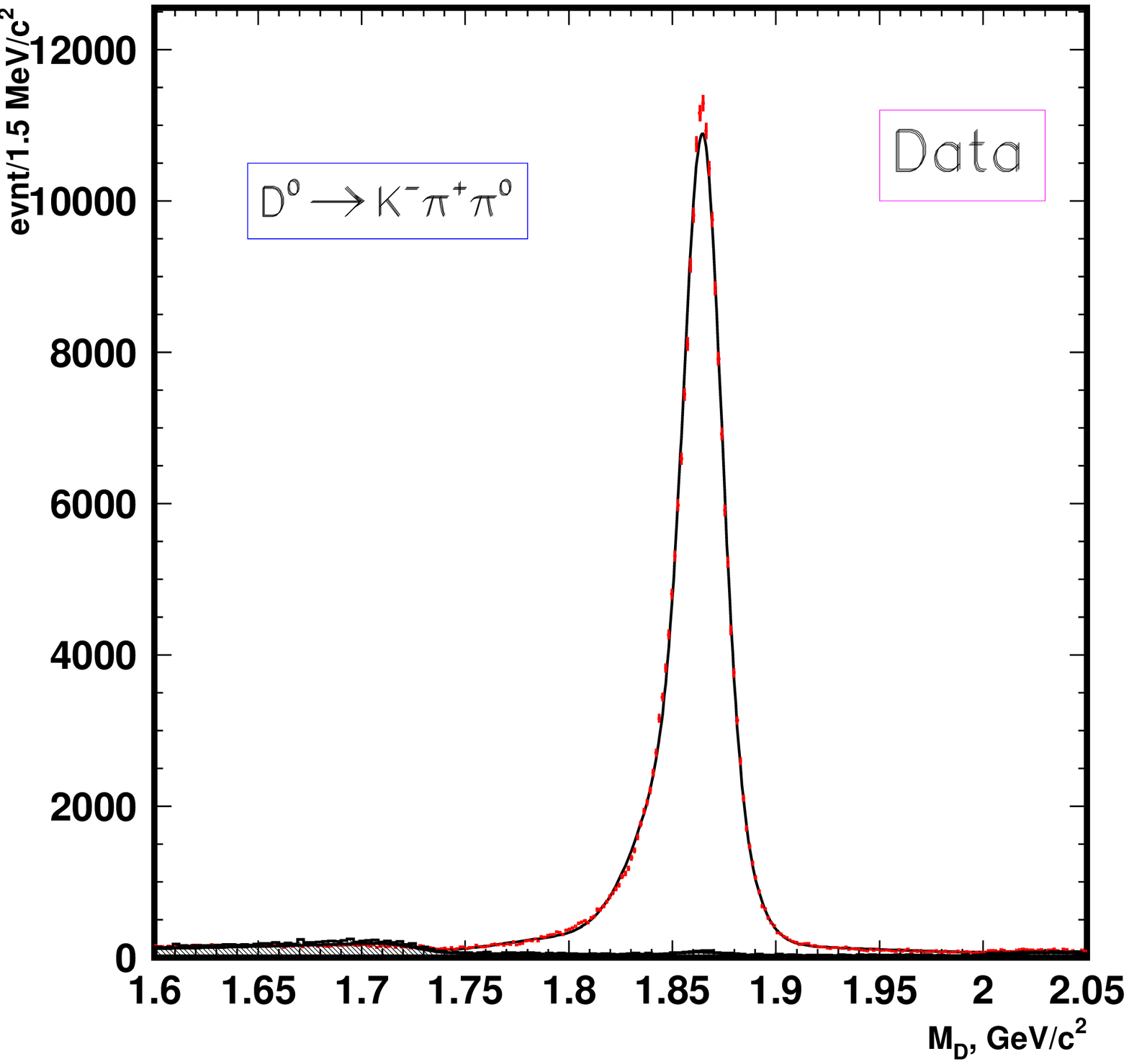}\hspace*{0.3in}
\includegraphics[height=2.7in]{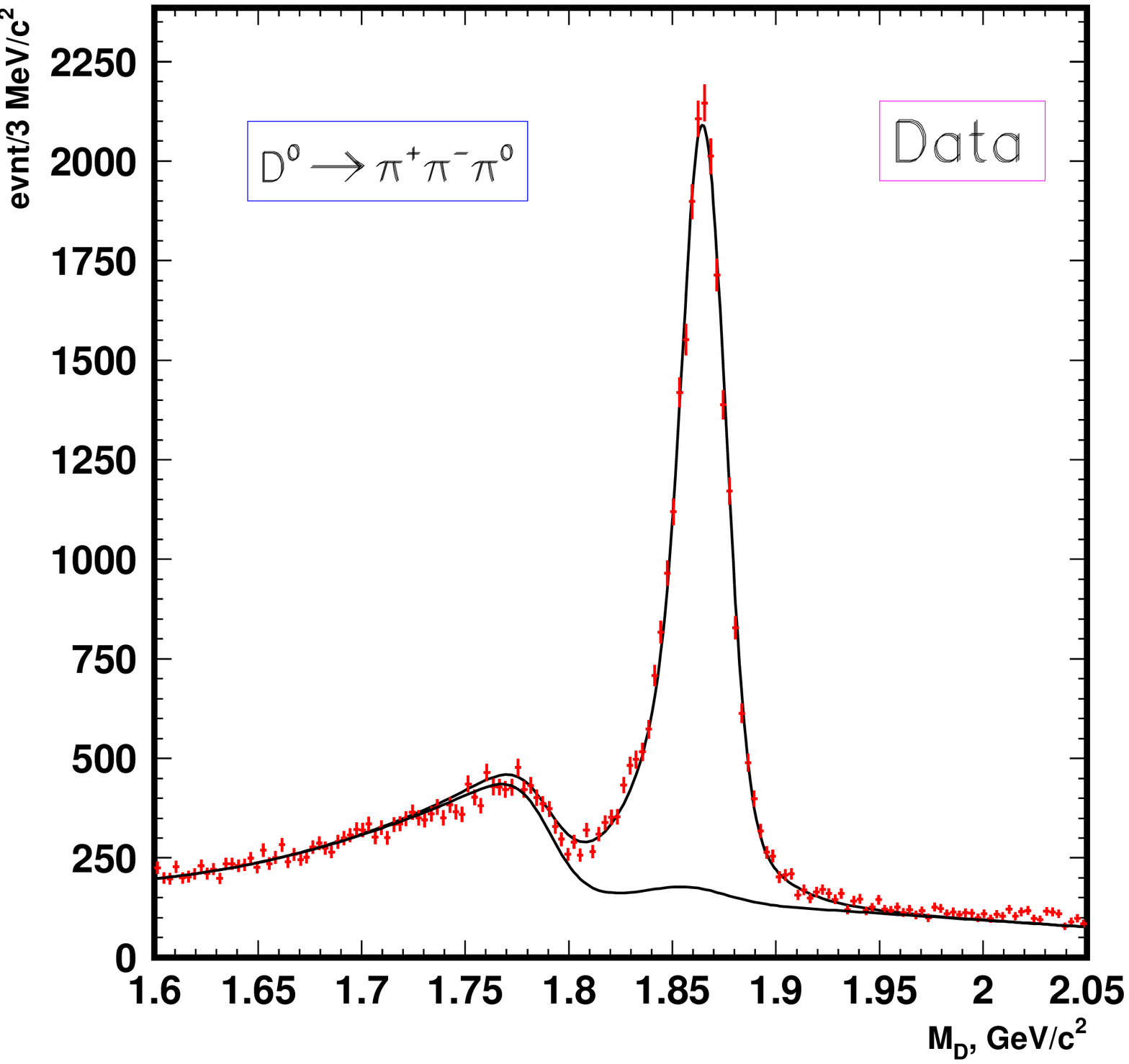}
\caption{Belle data for (left) $\Dz\to\Km\pip\piz$ decays and (right) $\Dz\to\pip\pim\piz$ decays.}
\label{fig:Belle-Dzpipipi}
\end{center}
\end{figure}

BaBar~\cite{babar:scsd} and Belle~\cite{belle:scsd} have used their enormous data samples to measure precisely the ratio $\calB(\Dz\to\pip\pim\piz)/\calB(\Dz\to\Km\pip\piz)$ of a singly-Cabibbo-suppressed hadronic decay to a similar Cabibbo-favored hadronic decay.  The BaBar and Belle data are illustrated in Figures~\ref{fig:BaBar-Dzpipipi} and \ref{fig:Belle-Dzpipipi}, respectively.  The invariant masses of $D$ candidates are used in these analyses, since these data were taken at the $\Upsilon(4S)$ where many $D\Dbar$ channels with multipions are open (rather than only the $\Dz\Dzbar$ and $\Dp\Dm$ channels at the $\psidprime$) so beam-constrained masses cannot be used.   Therefore the mass resolutions are substantially worse in the BaBar and Belle data, than in the corresponding CLEO-c data (compare Figures~\ref{fig:BaBar-Dzpipipi} and \ref{fig:Belle-Dzpipipi} with \Fig{fig:Mpippimpiz}.  However, signal to background ratios in the BaBar and Belle data are substantially better than the corresponding ratio in the CLEO-c data.  The results of the three measurements are given in \Tab{tab:BBC-scsd}.  Some of the strengths and weaknesses of these measurements are apparent from the yields, errors, and luminosities in the table.  Note that the Cabibbo suppression of $\calB(\Dz\to\pip\pim\piz)$ is about a factor of 10 relative to $\calB(\Dz\to\Km\pip\piz)$, which is the $\Dz$ hadronic decay mode with the largest branching fraction.

\begin{table}[htb]
\Begtabular{l|rcc}
& Yield & $\calB(\Dz\to\pip\pim\piz)/\calB(\Dz\to\Km\pip\piz)$ & Luminosity  \\ \hline
BaBar  & $60,426 \pm 343$ & $(10.59  \pm 0.06 \pm 0.13)\times 10^{-2}$ & 232~\fbinv\ \\ 
Belle  & $22,803 \pm 203$ & $(~9.71 \pm 0.09 \pm 0.30)\times 10^{-2}$  & 357~\fbinv\ \\
CLEO-c & $10,834 \pm 164$ & $(~9.01 \pm 0.18 \pm 0.39)\times 10^{-2}$ & ~281~\pbinv \\ \hline \hline
& & $\calB(\Dz\to\Km\Kp\piz)/\calB(\Dz\to\Km\pip\piz)$ \\ \hline
BaBar  & $10,773 \pm 122$ & $(2.37 \pm 0.03 \pm 0.04)\times 10^{-2}$ & 232~\fbinv\ \\
CLEO II & $~~~~151 \pm ~~42$ & $(0.95 \pm 0.26)\times 10^{-2}$ & ~2.7~\fbinv\ \\
\Endtabular
\caption{Measurements of SCSDs from BaBar, Belle, and CLEO.  The Belle result is preliminary, while the BaBar and CLEO results have been published.  The CLEO-c branching ratio is calculated from measured ratios of both $\calB(\Dz\to\pip\pim\piz)$ and $\calB(\Dz\to\Km\pip\piz)$ to $\calB(\Dz\to\Km\pip)$, without taking correlations into account.  Proper use of correlations would reduce the systematic error. \label{tab:BBC-scsd}}
\end{table}

\begin{figure}[htb]
\begin{center}
\includegraphics[width=2.7in]{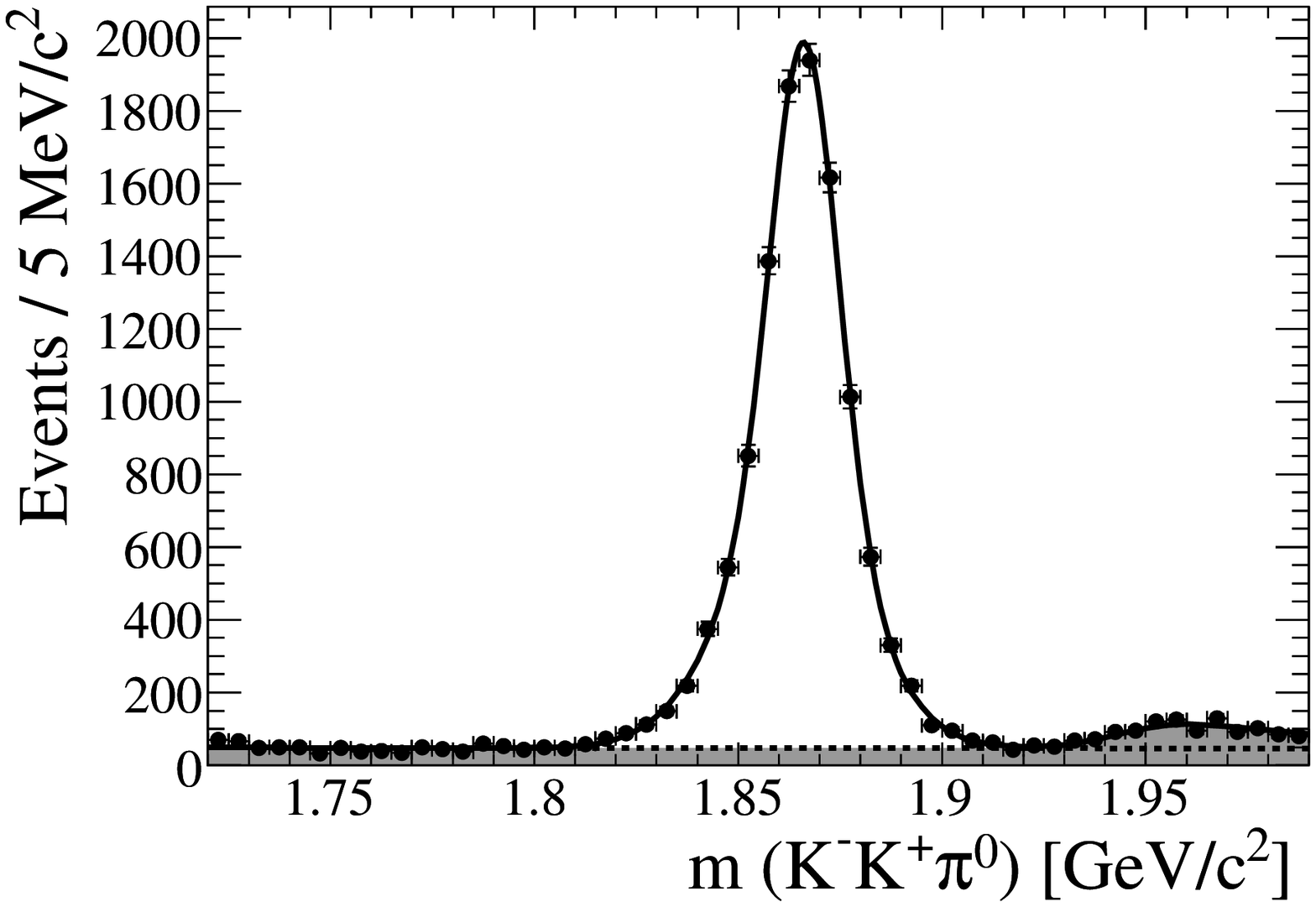}
\caption{BaBar data for $\Dz\to\Km\Kp\piz$ decays.~\cite{aps:babar:scsd,aps:extra}}
\label{fig:BaBar-DzKmKppiz}
\end{center}
\end{figure}

In the same analysis BaBar measured $\calB(\Dz\to\Km\Kp\piz)/\calB(\Dz\to\Km\pip\piz)$.  The data are illustrated in \Fig{fig:BaBar-DzKmKppiz} and the result is included in \Tab{tab:BBC-scsd}.  The BaBar result is significantly larger and much more precise than a previous CLEO result from a much smaller data sample~\cite{cleo:DKKX}. 

\subsection{Multi-Kaon Modes from FOCUS}

\begin{figure}[htb]
\begin{center}
\includegraphics[height=2.5in]{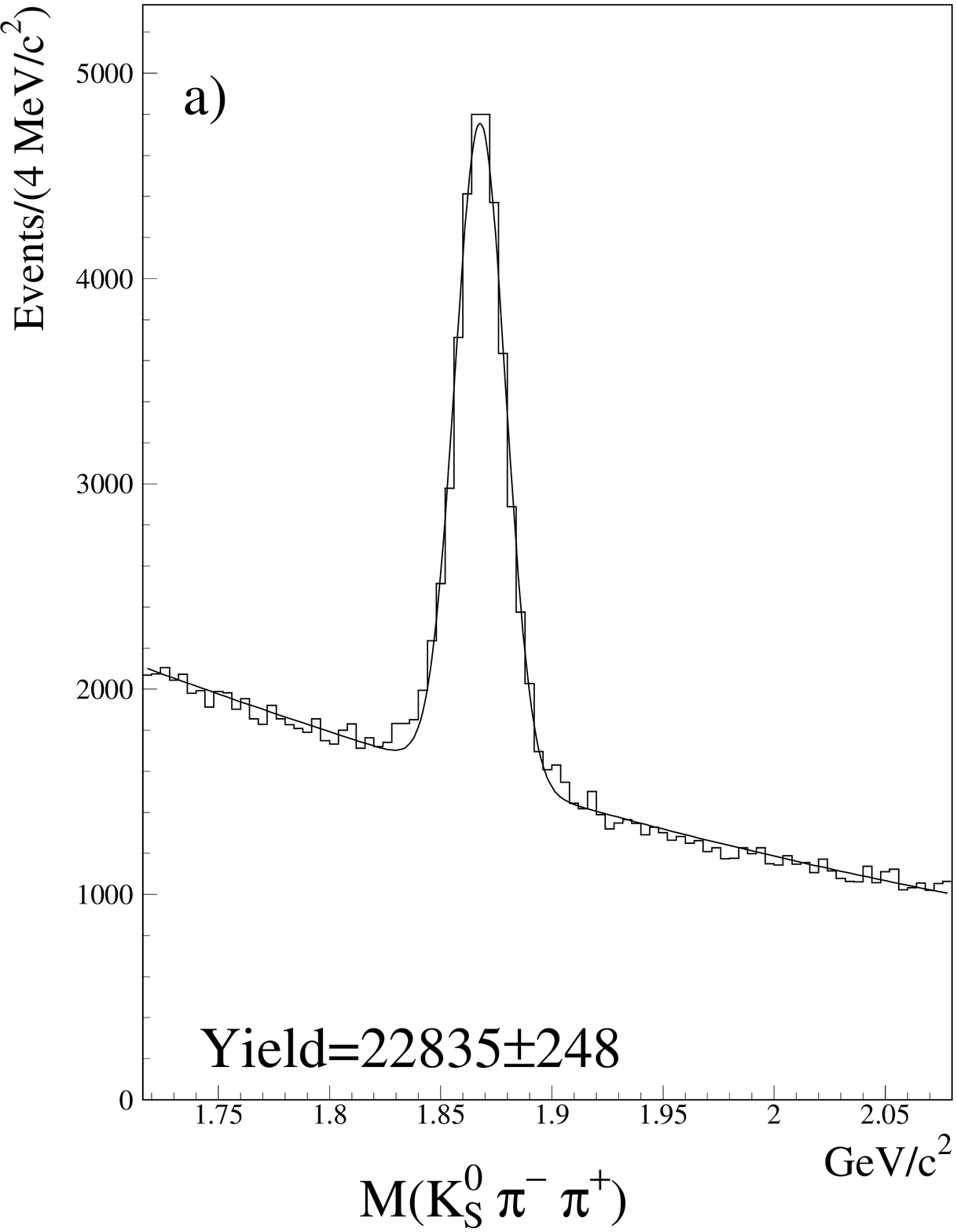}\hfill
\includegraphics[height=2.5in]{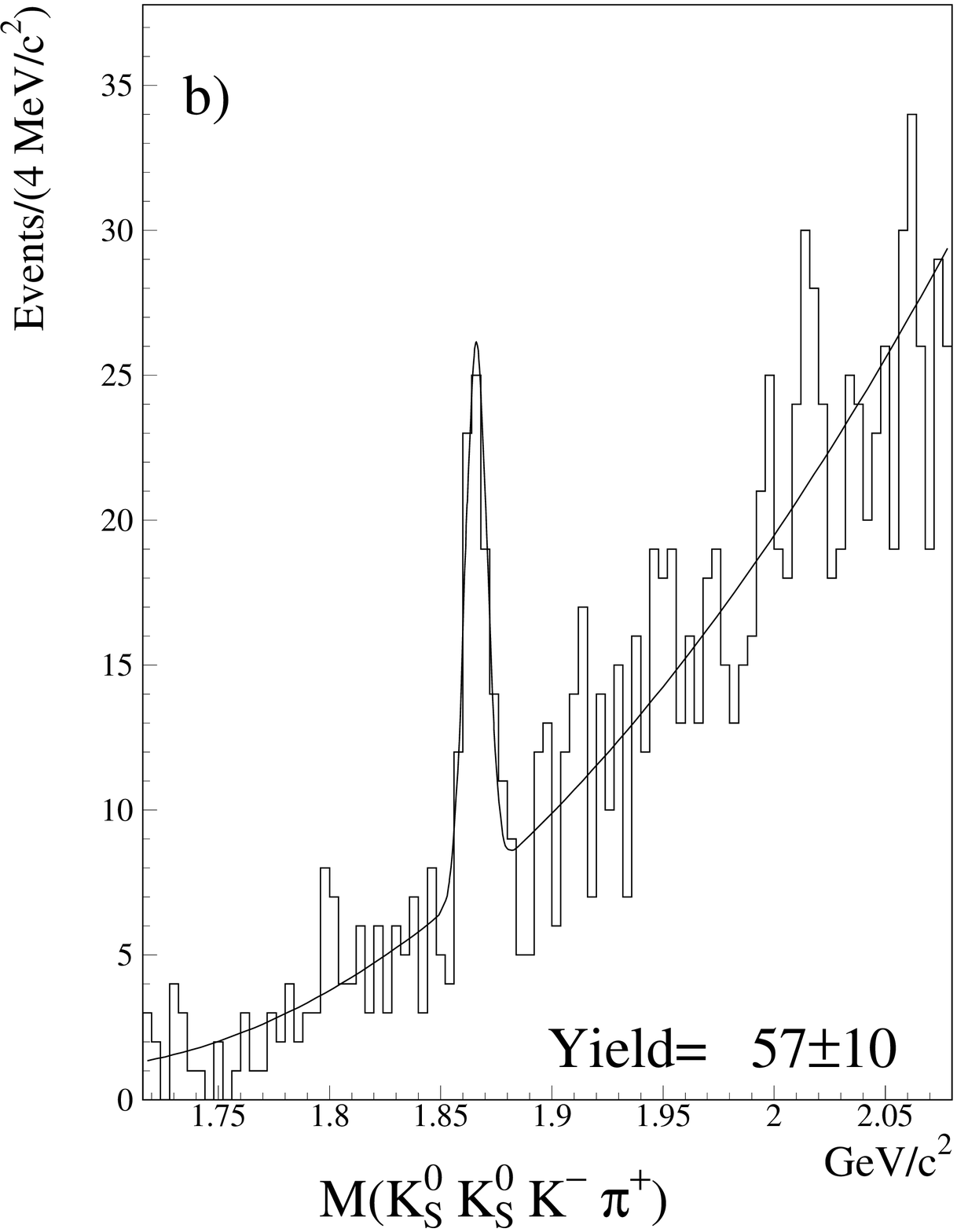}\hfill
\includegraphics[height=2.5in]{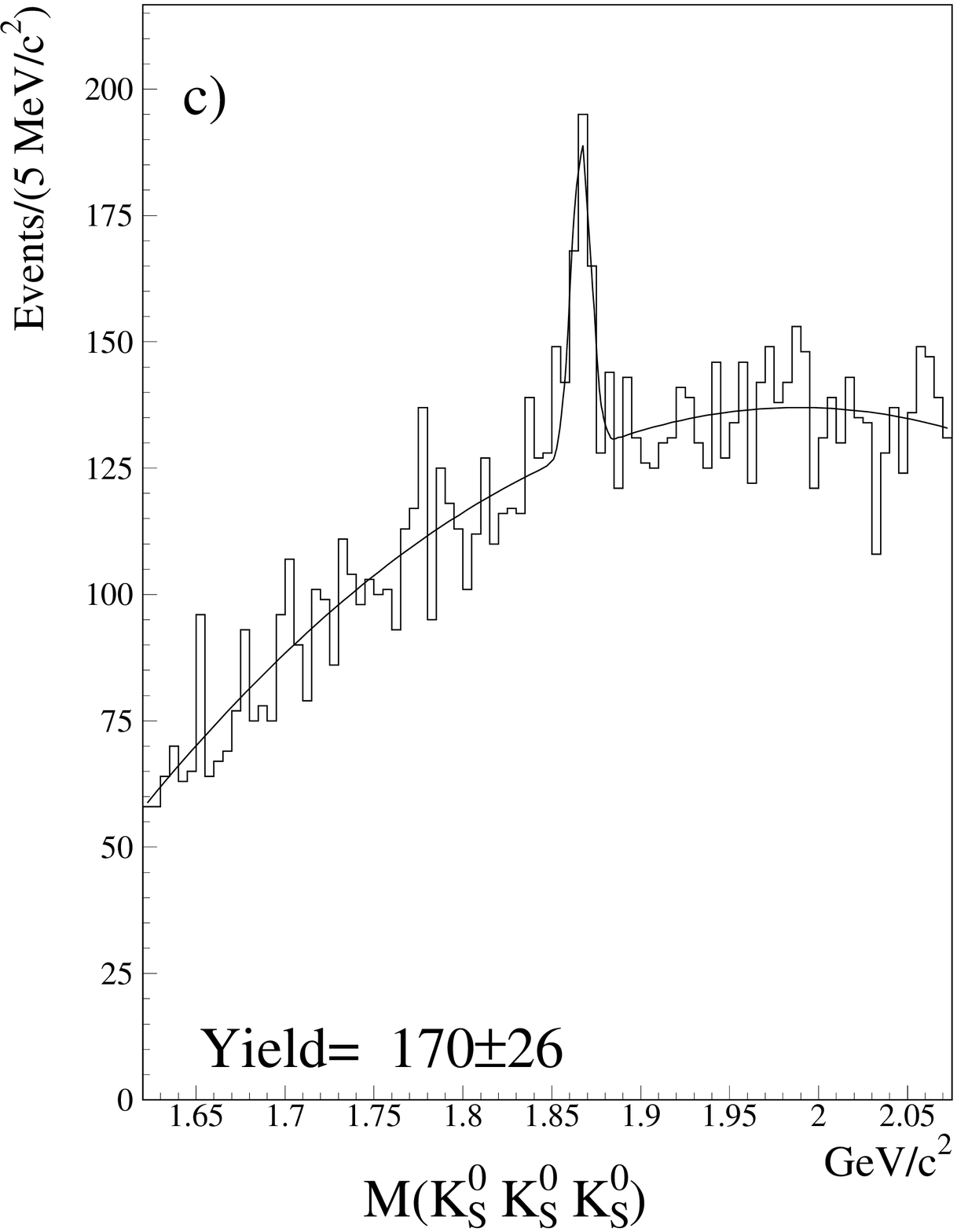}
\caption{FOCUS data for (left) $\Dz\to\KS\pip\pim$ decays, (middle) $\Dz\to\KS\KS K^\pm \pi^\mp$ decays, and (right) $\Dz\to\KS\KS\KS$ decays~\cite{plb:focus:multik}.}
\label{fig:FOCUS-SCSD}
\end{center}
\end{figure}

FOCUS studied $\Dz$ decays to final states with two or three charged or neutral kaons~\cite{focus:multik,focus:Dalitz-DzKpKmpippim}.  The data for the reference mode $\Dz\to\KS\pip\pim$ and two of the neutral kaon signal modes are illustrated in \Fig{fig:FOCUS-SCSD} and the ratios of signal branching fractions to the reference branching fraction are given in \Tab{tab:FOCUS-SCSD}.  This is the first observation of the modes $\Dz\to\KS\KS K^\pm\pi^\mp$, which are combined to one branching fraction.  The table also includes a measurement of the branching ratio $\calB(\Dz\to\Km\Kp\pip\pim)/\calB(\Dz\to\Km\pip\pip\pim)$.  This branching ratio was measured in a separate FOCUS analysis of resonant substructure in the the four-body decay $\Dz\to\Km\Kp\pip\pim$ described in \Sec{sec:focus-Dalitz-KKpipi}.

\begin{table}[htb]
\begin{center}
\Begtabular{l|cc}
Mode \hspace*{6em} & FOCUS $\calB_\textrm{mode}/\calB_\textrm{ref}$ (\%) \\[2pt] \hline
$\Dz\to\Kz\Kzbar$ & $1.44 \pm 0.32 \pm 0.16$ & ~~SCSD~~ \\
$\Dz\to\KS\KS\pip\pim$ & $2.08 \pm 0.35 \pm 0.21$ & SCSD \\
$\Dz\to\KS\KS K^\pm \pi^\mp$ & $1.06 \pm 0.19 \pm 0.10$ & CF \\
$\Dz\to\KS\KS\KS$ & $1.79 \pm 0.27 \pm 0.26$ & CF \\ \hline
$\Dz\to\Km\Kp\pip\pim$ & $2.95 \pm 0.11 \pm 0.08$ & SCSD \\
\Endtabular
\end{center}
\caption{FOCUS branching fractions for Cabibbo favored (CF) decays and singly-Cabibbo suppressed $\Dz$ decays to multiple kaons.  For the neutral kaon modes, the reference branching fraction is
$\calB(\Dz\to\Kzbar\pip\pim)$, while the reference branching fraction for the charged kaon mode is $\calB(\Dz\to\Km\pip\pip\pim)$.}
\label{tab:FOCUS-SCSD}
\end{table}

From the table it is evident that Cabibbo-suppression, relative to the Cabibbo-favored reference mode, is on the order of a factor of 100 for the two-kaon decay modes.  However, the Cabibbo-favored decays to three kaons are suppressed by comparable factors.  Presumably this is due to the necessity of popping an $s\sbar$ quark-antiquark pair to produce the additional two kaons.     

\subsection{\boldmath The Doubly-Cabibbo-Suppressed Decay $\Dp\to\Kp\pim$ \label{sec:DpKppim}}

\begin{figure}[htb]
\begin{center}
\includegraphics[width=1.9in]{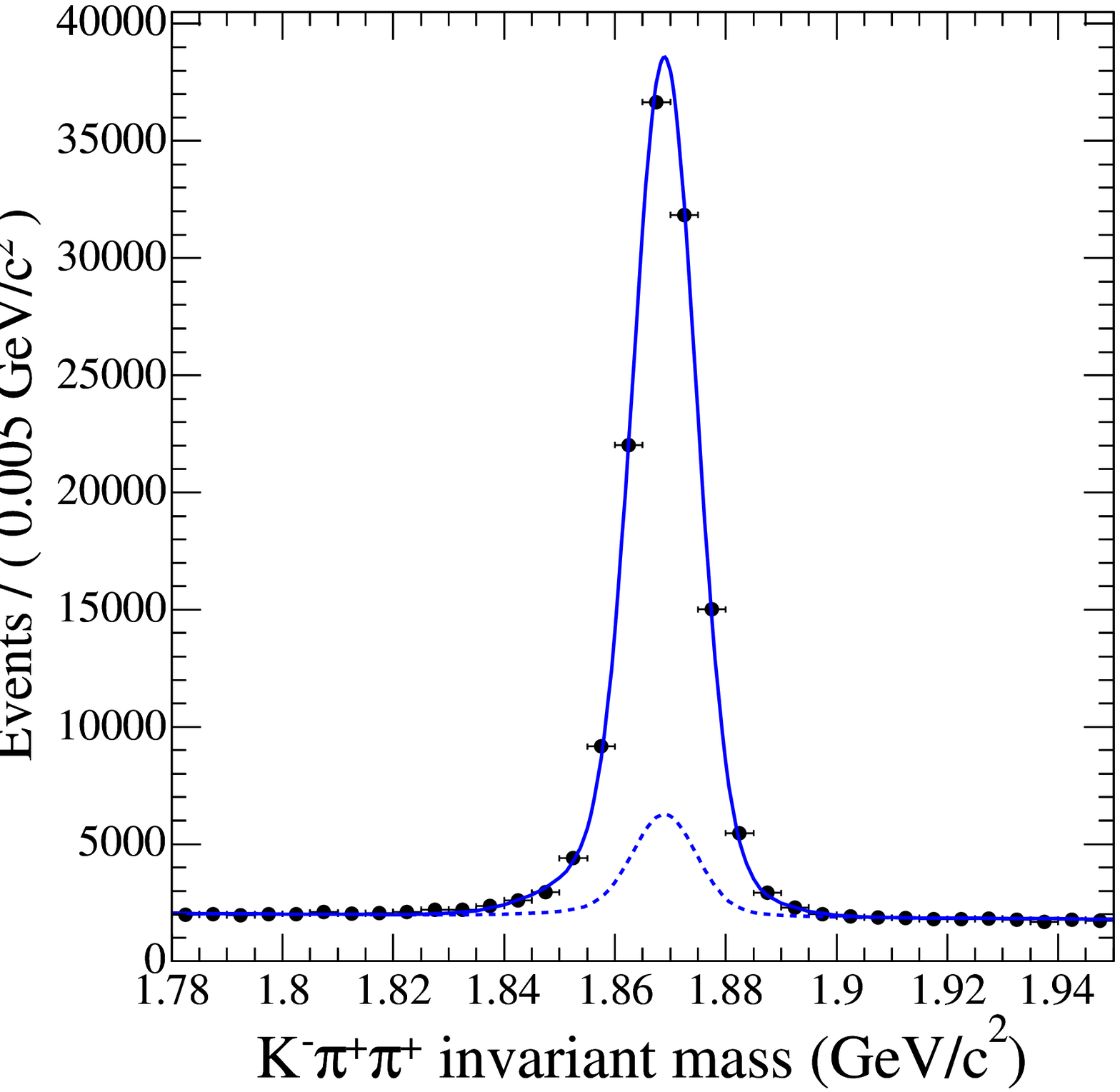}\hfill
\includegraphics[width=1.9in]{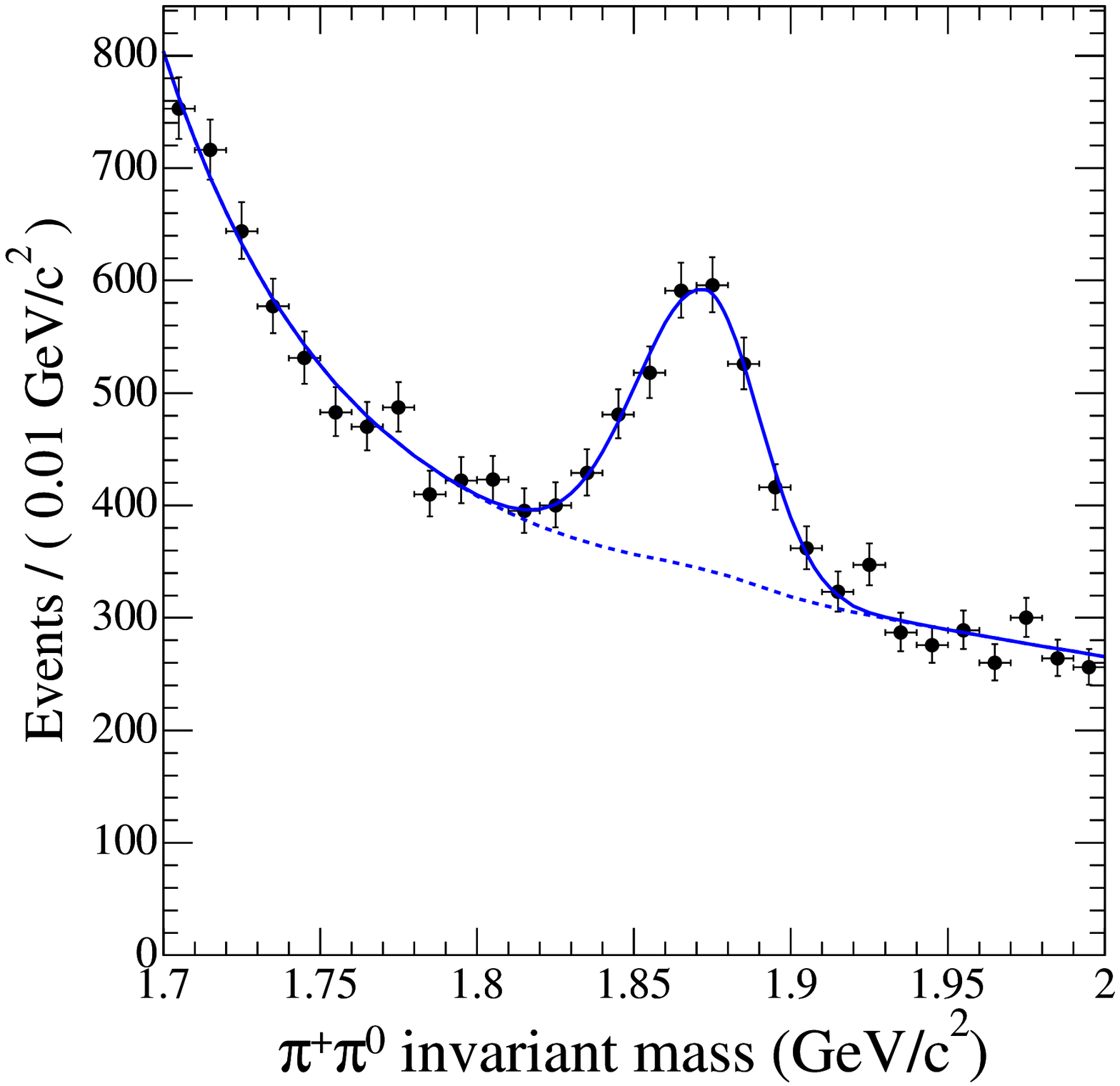}\hfill
\includegraphics[width=1.9in]{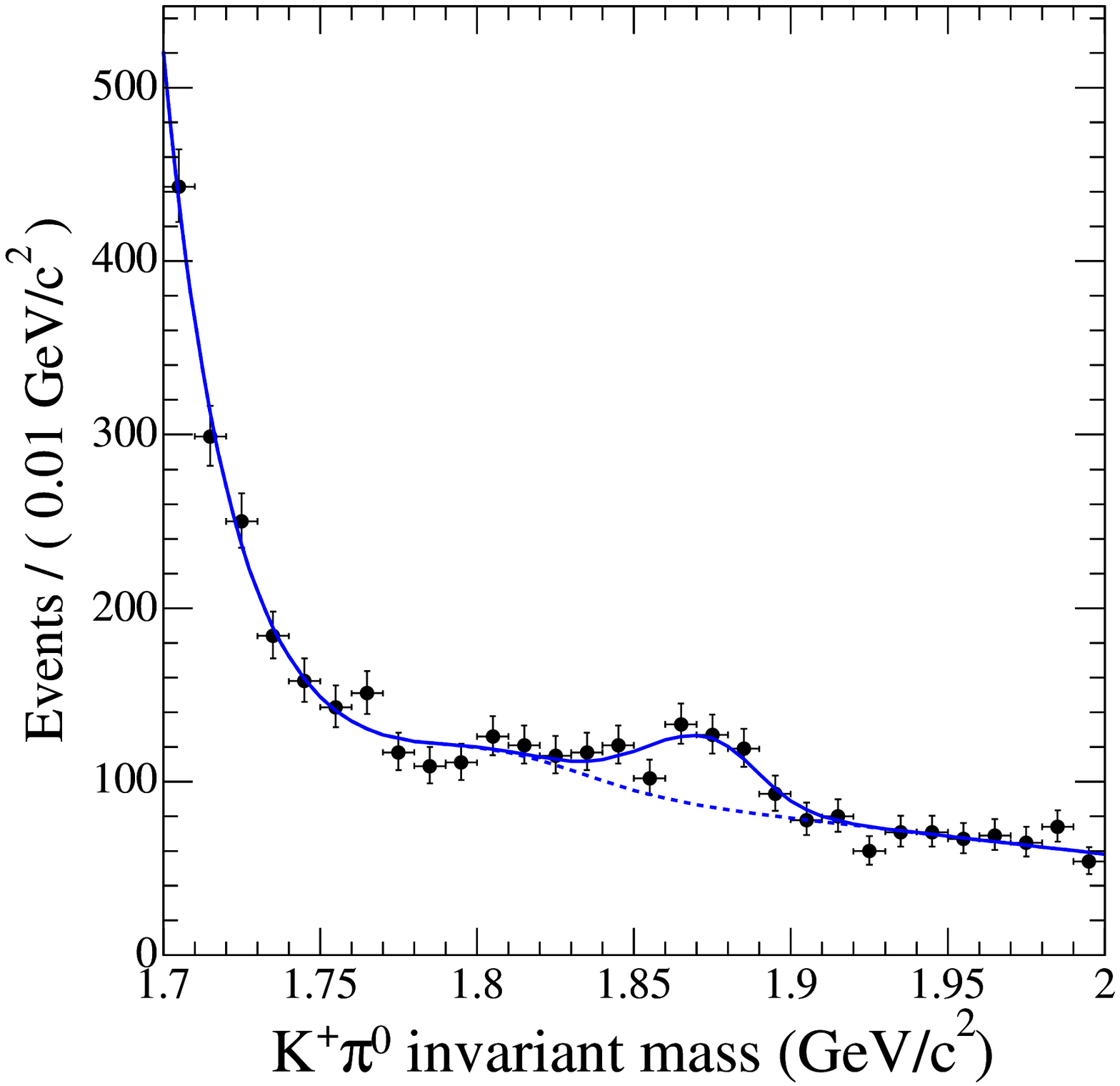}
\caption{BaBar data for (left) $\Dp\to\Km\pip\pip$ decays, (middle) $\Dp\to\pip\piz$ decays, and (right) $\Dp\to\Kp\pim$ decays.~\cite{aps:extra,aps:babar:dcsd}}
\label{fig:BaBar-DCSD}
\end{center}
\end{figure}

\begin{figure}[htb]
\begin{center}
\includegraphics[width=5.9in]{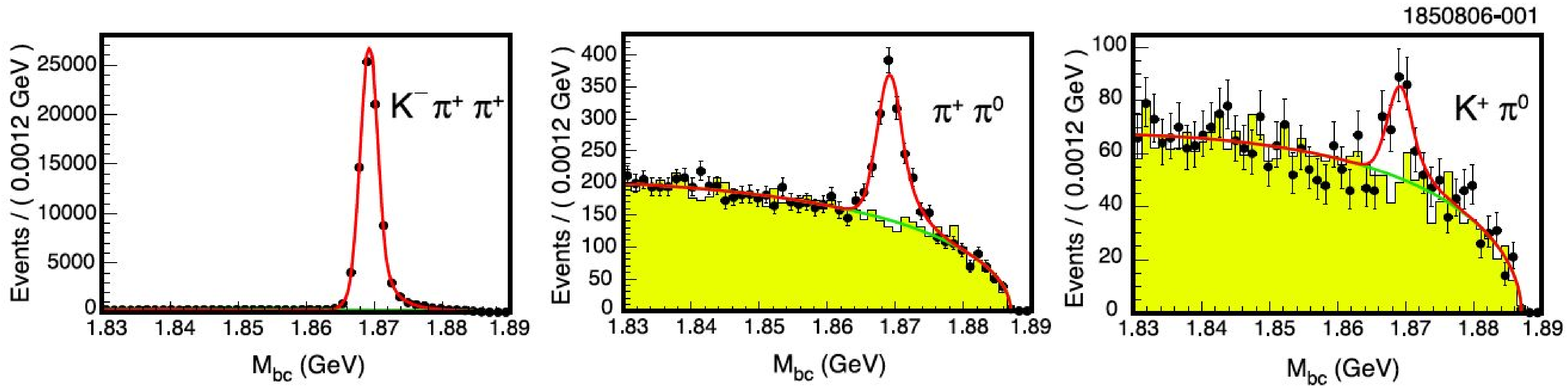}\hfill
\caption{CLEO-c data for (left) $\Dp\to\Km\pip\pip$ decays, (middle) $\Dp\to\pip\piz$ decays, and (right) $\Dp\to\Kp\pim$ decays.}
\label{fig:cleo-dcsd}
\end{center}
\end{figure}

BaBar has reported the first observation of the DCSD $\Dp\to\Kp\piz$~\cite{babar:dcsd} and the precise measurement of a branching ratio for the SCSD decay $\Dp\to\pip\piz$ mentioned in \Sec{sec:Dnpi}.  The BaBar data are illustrated in \Fig{fig:BaBar-DCSD}, and the CLEO data~\cite{cleo:dcsd} for the same modes are illustrated in \Fig{fig:cleo-dcsd}.  Note that the mass range of the CLEO data is much narrower than the mass range of the BaBar data, illustrating the improvements in mass resolution gained by being able to use beam-constrained masses instead of invariant mass.  The ratios of the signal branching fractions $\calB(\Dp\to\pip\piz)$ and $\calB(\Dp\to\Kp\piz)$ of the signal modes to reference branching fraction $\calB(\Dp\to\Km\pip\pip)$ are given in \Tab{tab:DpKppiz}.  BaBar's ability to use tighter cuts on their enormous data sample to compensate for the better mass resolution of the CLEO-c data is evident from the figures and results.  The Cabibbo suppression of the $\Dp\to\Kp\piz$ mode is close to a factor of 5. 

\begin{table}[htb]
\begin{center}
\Begtabular{l|cc}
 & $\Dp\to\pip\piz$~ (SCSD) & $\Dp\to\Kp\piz$~ (DCSD) \\
 & $\calB_\textrm{mode}/\calB_\textrm{ref}$~ & 
                $\calB_\textrm{mode}/\calB_\textrm{ref}$~ \\[2pt] \hline
BaBar    & ~$(1.33 \pm 0.11 \pm 0.09)\times 10^{-2}$~ & 
           ~$(2.68 \pm 0.50 \pm 0.26)\times 10^{-3}$~ \\
CLEO-c~~ &  $(1.33 \pm 0.07 \pm 0.06)\times 10^{-2}$ & 
            $(2.40 \pm 0.38 \pm 0.16)\times 10^{-3}$ \\
\Endtabular
\caption{The ratios of the branching factions for the SCSD decay $\Dp\to\pip\piz$ and the DCSD decay $\Dp\to\Kp\piz$ from BaBar and CLEO-c, to the branching fraction for the reference mode $\Dp\to\Km\pip\pip$. The CLEO-c result for $\Dp\to\pip\piz$ is from the SCSD analysis in \Sec{sec:Dnpi}.\label{tab:DpKppiz}}
\end{center}
\end{table}

\subsection{\boldmath Comparison of $D\to\KS\pi$ and $D\to\KL\pi$ Decay Rates}

\begin{figure}[htb]
\begin{center}
\includegraphics[height=2.0in]{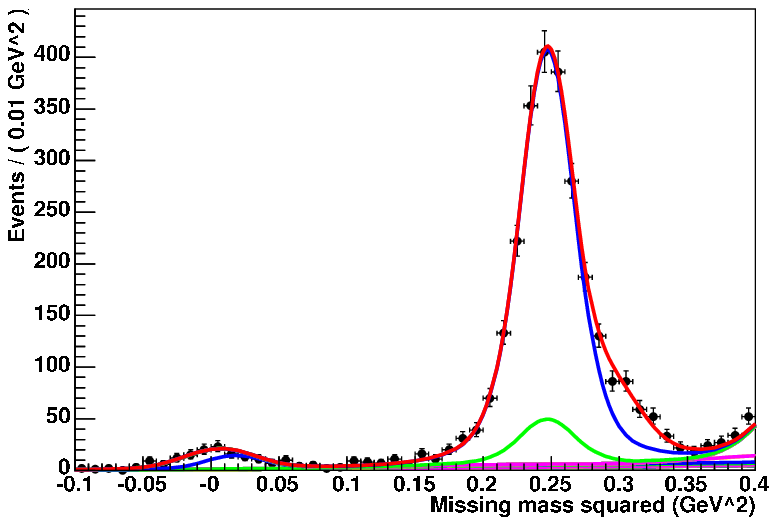}\hfill
\includegraphics[height=2.0in]{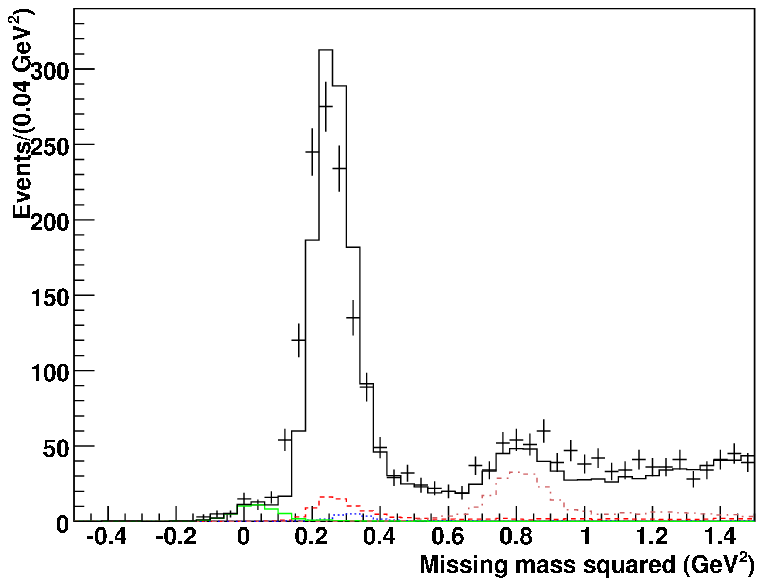}
\caption{CLEO data for (left) $\Dz\KL\piz$ decays and (right) $\Dp\to\KL\pip$ decays.}
\label{fig:DKLpi}
\end{center}
\end{figure}

Cabibbo-Favored and Doubly-Cabibbo-Suppressed amplitudes contribute to the decay $D\to\Kz\pi$.
The observed final states are $D\to\KS\pi$ and $D\to\KL\pi$, and Bigi and Yamamoto pointed out that interference between CF and DCS amplitudes can lead to different rates for $D\to\KS\pi$ and $D\to\KL\pi$~\cite{Bigi:Yamamoto}.  CLEO has measured preliminary branching fractions for these decays by fully reconstructing $D\to\KS\pi$ decays and reconstructing $D\to\KL\pi$ decays using missing masses~\cite{cleo:DKzpi}.   The CLEO data for $D\to\KL\pi$ decays are illustrated in \Fig{fig:DKLpi}.  The preliminary results given in \Tab{tab:DKzpi} are reported in terms of the ratio $R(D)$ defined by:  
\[ R(D) \equiv {{\calB(D\to\KS\pi) - \calB(D\to\KL\pi)} \over
                {\calB(D\to\KS\pi) + \calB(D\to\KL\pi)}} \]
The measured value of $R(\Dp)$ is consistent with zero, while $R(\Dz)$ is significantly larger than zero.  U-spin and SU(3) predict~\cite{Rosner:Dkzpi} $R(\Dz) = 2 \tan^2(\theta_c)$ giving $R(\Dz) = 0.109 \pm 0.001$, which is in good agreement with the experimental result.  On the other hand, $R(\Dp)$ is not so simple because internal spectator diagrams contribute to both 
$\Dp\to\Kzbar\pip$ and $\Dp\to\Kz\pip$, but external diagrams contribute to the former and annihilation diagrams contribute to the latter. 

\begin{table}[htb]
\begin{center}
\Begtabular{l|c}
$R(\Dp)$ & $0.030 \pm 0.023 \pm 0.025$ \\
$R(\Dz)$ & $0.122 \pm 0.024 \pm 0.030$ \\
\Endtabular
\caption{CLEO-c Preliminary measurements of the normalized branching fraction differences between $D\to\KL\pi$ decays and  $D\to\KS\pi$ decays.}
\label{tab:DKzpi}
\end{center}
\end{table}

\section{\boldmath Absolute $\Ds$ Branching Fractions}

Previously, absolute $\Ds$ branching fractions have not been well known, primarily due to difficulties of determining the total number of $\Ds$ mesons produced in the experiment~\cite{cleo:dsabsolute-old,babar:dsabsolute}.  CLEO has now addressed this problem with a double tag measurement of absolute $\Ds$ branching fractions in $\elp\elm$ collisions just above the $\Dspm\Dsmpstar$ threshold~\cite{cleo:dsabsolute}.

\subsection{\boldmath The $\Ds$ Production Cross Section}

\begin{figure}[htb]
\begin{center}
\includegraphics[width=2.9in]{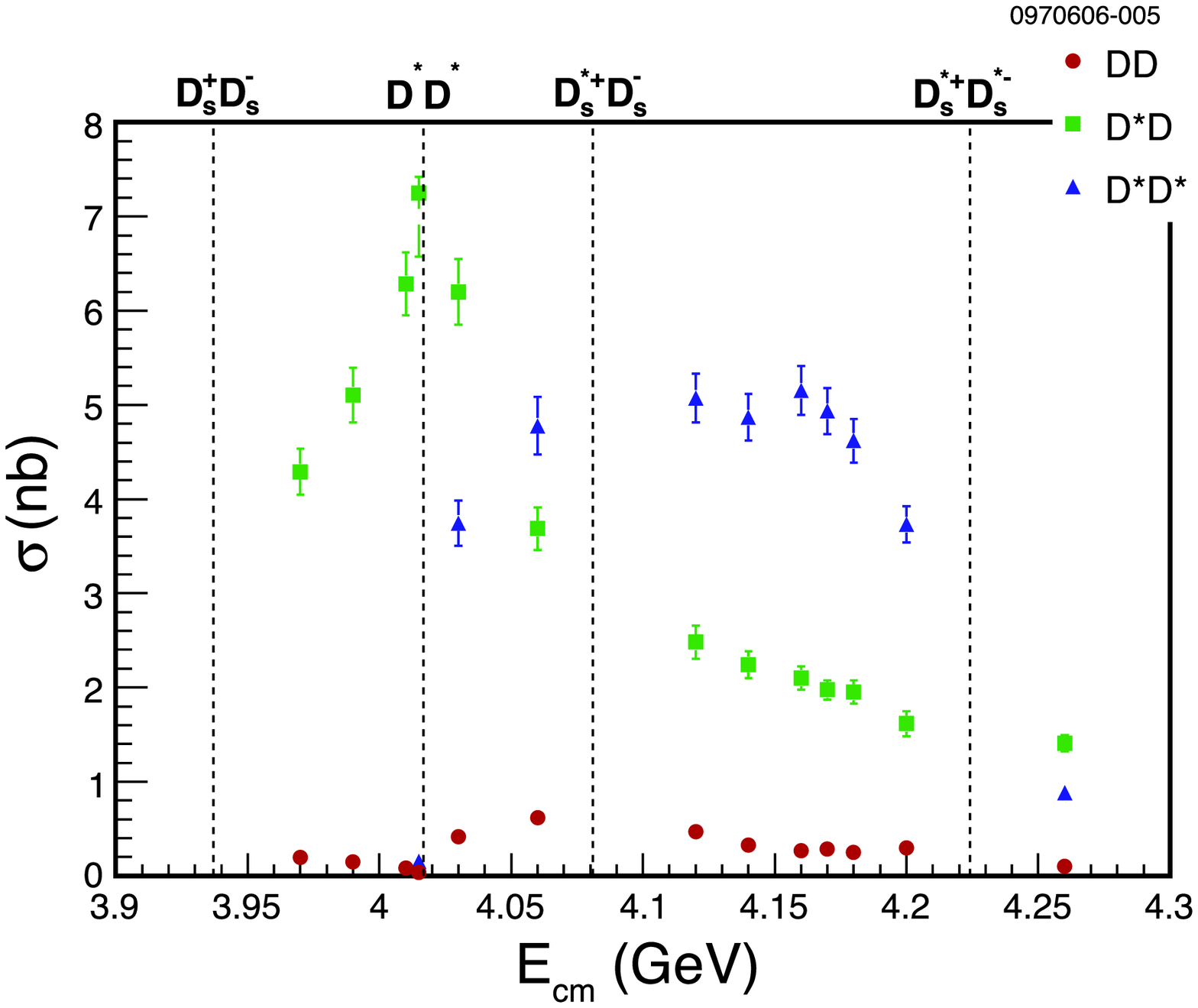}\hfill
\includegraphics[width=3.0in]{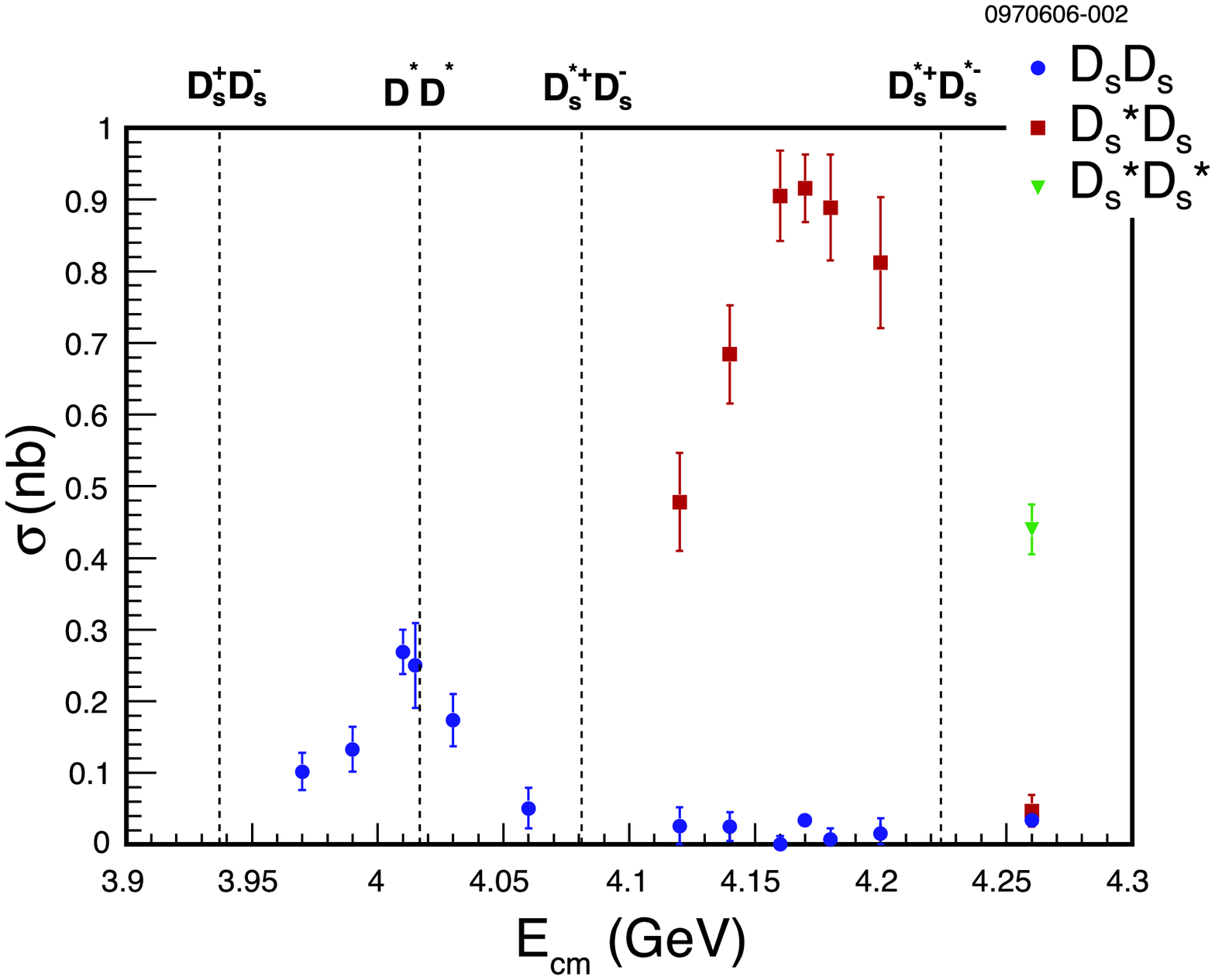}
\caption{Preliminary CLEO-c cross sections for production of (left) $D^{(*)}\bar{D}^{(*)}$ and (right) $D_s^{+(*)}D_s^{-(*)}$ above the threshold for $D\Dbar$ production in $\elp\elm$ annihilation.  In each graph thresholds are indicated by the vertical dashed lines.}
\label{fig:sigDDbar}
\end{center}
\end{figure}

Although measurements of the $\elp \elm$ annihilation cross section, $\sigma(\elp \elm)$, at energies above $E_{cm} = 3.8$~GeV have existed for some time~\cite{pdg:06}, little was known about the composition of the final states.  In order to find a favorable point for studying DT $\Dspm\Dsmpbar$ events, CLEO scanned that region with integrated luminosities of $\sim 5$~\pbinv\ per point and fast turnaround of results for feedback.  After the scan, more luminosity was accumulated at or near $E_{cm} = 4.17$~GeV where the cross section for $\Dspm\Dsmpstar$ production peaks with $\sigma(\elp \elm \to \Dspm\Dsmpstar) \approx 0.9$~nb.  Preliminary measurements~\cite{cleo:dsabsolute} of cross sections for producing various $D\Dbar$, $D\Dstarbar$, and $\Dstar\Dstarbar$ pairs are illustrated in \Fig{fig:sigDDbar}.  CLEO used a total of 195~\pbinv\ of data at $E_{cm} = 4.17$~GeV in the analysis of $\Ds$ decays described below.   

\subsection{\bigskip Analysis of $\Dspm\Dsmpstar$ Events}

\begin{figure}[htb]
\begin{center}
\includegraphics[width=3.5in]{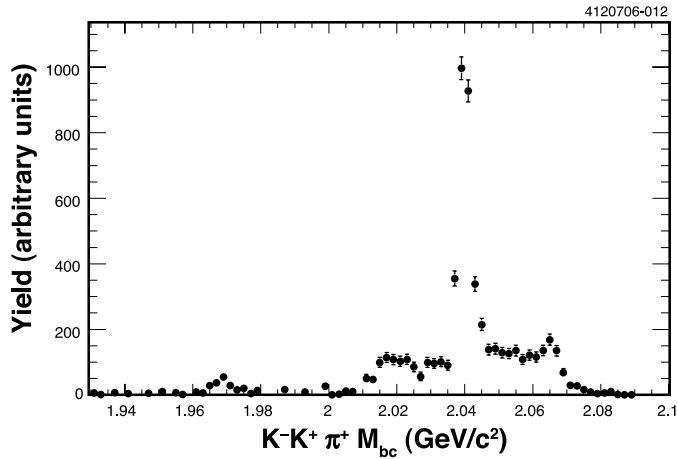}
\caption{Histogram of $\Mbc$ for $\Dsp\to\Km\Kp\pip$ events from CLEO-c data.  The narrow peak at the $\Mbc = 2.04$~GeV is from $\Dspm$ produced directly, while the broad peak from $\Mbc\approx 2.01$~GeV to $\Mbc\approx 2.07$~GeV is from $\Dspm$ produced in $\Dspmstar$ decay.}
\label{fig:DsMbc}
\end{center}
\end{figure}

CLEO chose to ignore the $\gamma$ or $\piz$ from $\Dsstar$ decay to avoid the low efficiency for detecting the low energy $\gamma$ or $\piz$ and the uncertainties in the efficiency for detecting either one of them.  Instead, CLEO selects $\Dspm\Dsmpstar$ events using the invariant mass $\MDs$ of the $\Ds$ candidates and their beam-constrained mass $\Mbc$, which is actually a proxy for momentum of the candidate.  The $\Mbc$ distribution for the $\Ds$ candidates that were produced directly in the annihilation is relatively narrow, while the $\Mbc$ distribution for $\Ds$ candidates resulting from $\Dsstar$ decay is much broader.  This is illustrated in \Fig{fig:DsMbc}.  Neither distribution is centered exactly on the $\Ds$ mass, because the beam energy is not exactly the energy of either the direct $\Ds$ or the average energy of the $\Ds$ from $\Dsstar$ decay.  

\begin{figure}[htb]
\begin{center}
\includegraphics[width=1.9in]{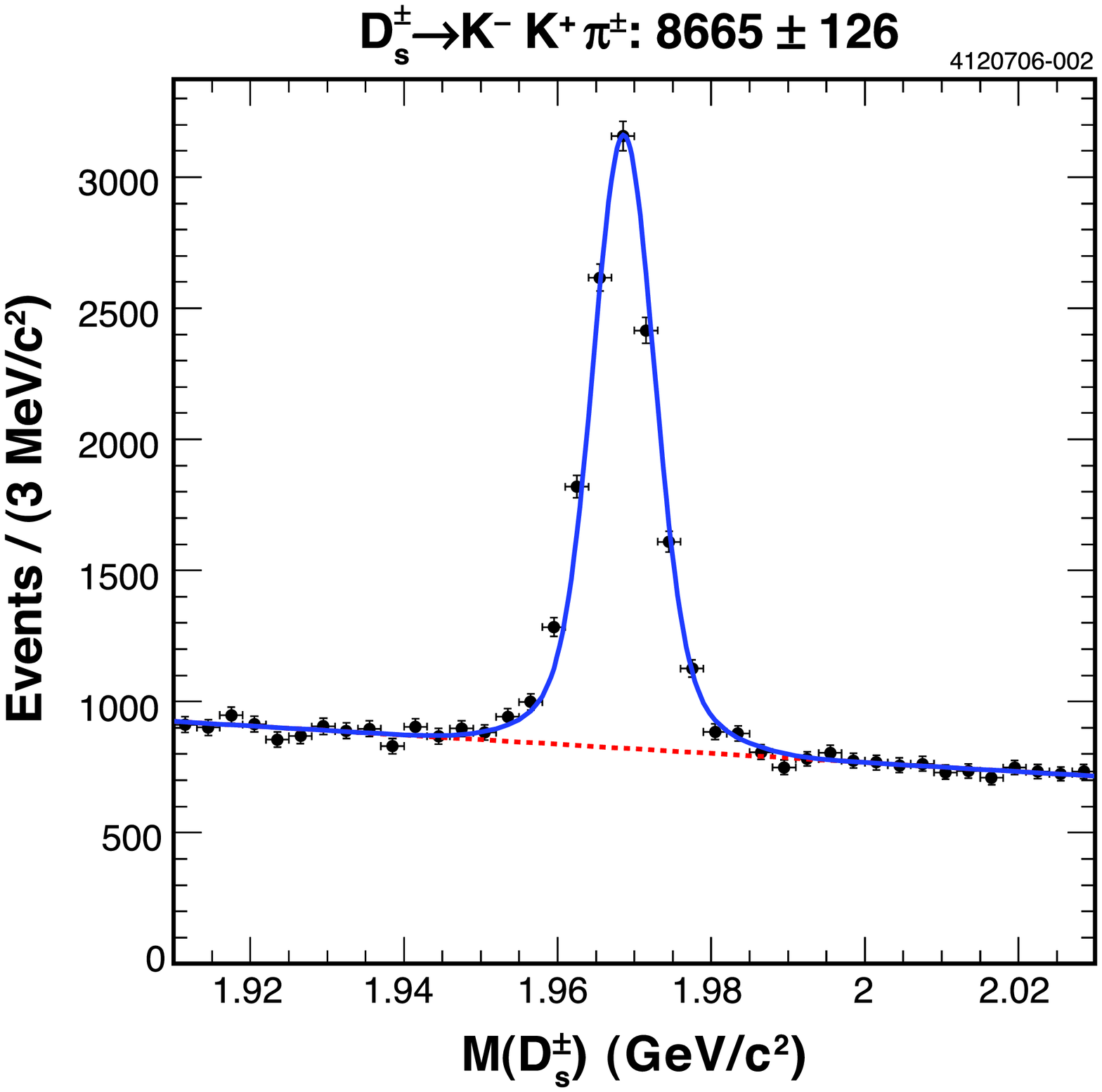}
\includegraphics[width=1.9in]{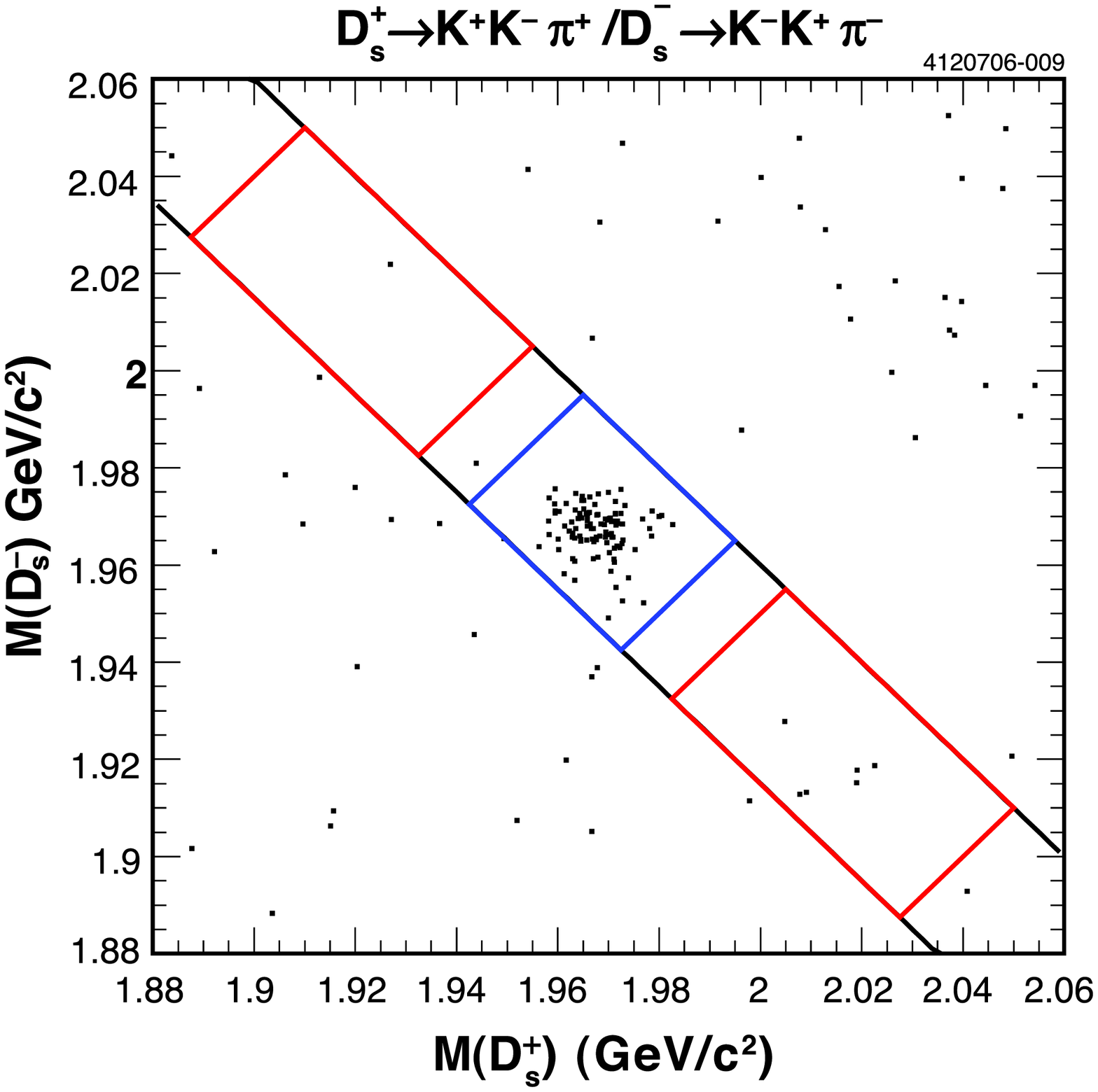}
\includegraphics[width=1.85in]{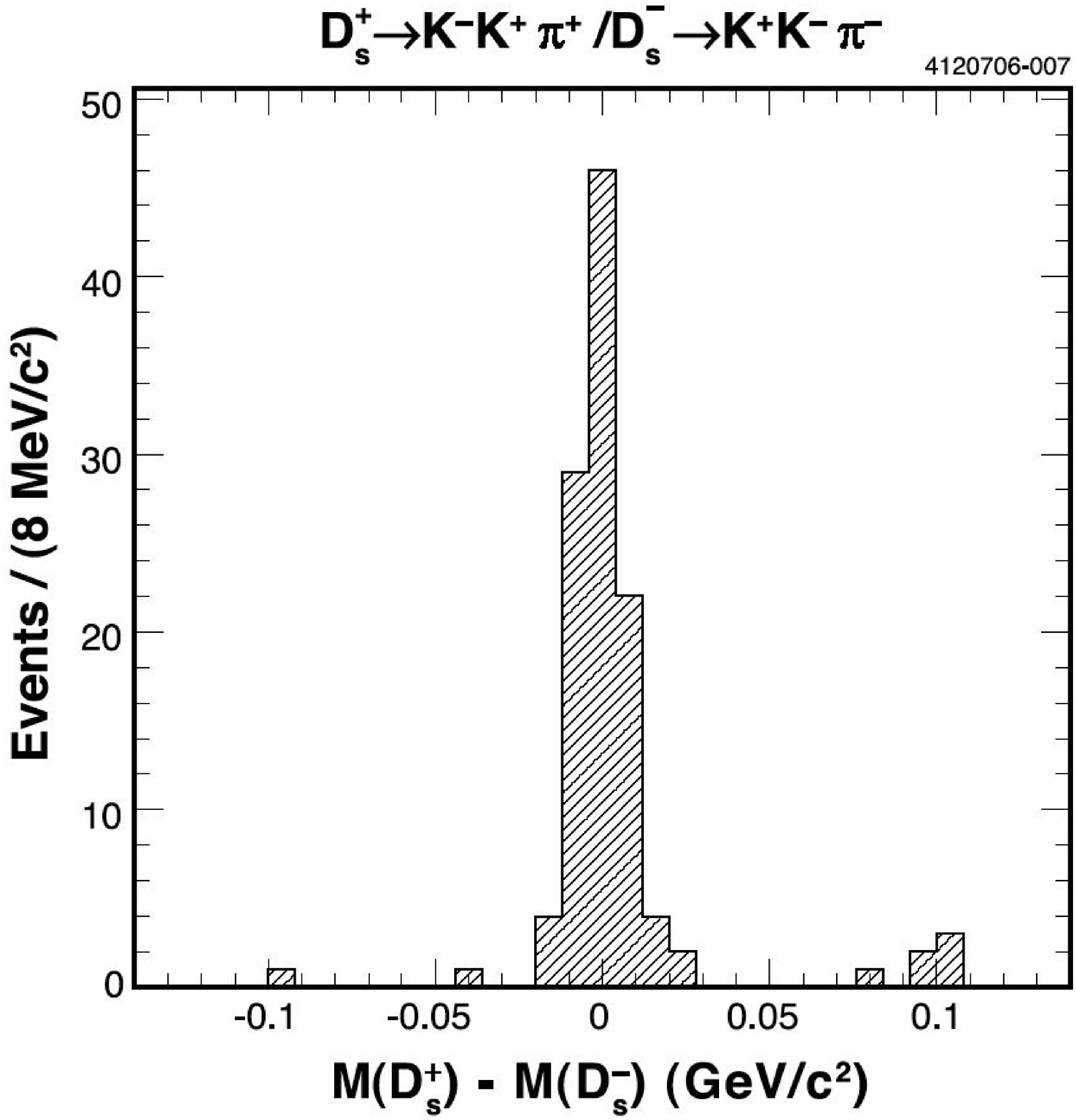}
\caption{Plots illustrating the CLEO-c $\Dspm$ decay data: (left) the invariant mass distribution for single-tag $\Dspm\to\Km\Kp\pipm$ candidates, (middle) two-dimensional histogram of $M(\Dsm)$ vs. $M(\Dsp)$ for double-tag $\Dspm\to\Km\Kp\pipm$ candidates, (right) the mass difference $M(\Dsp) - M(\Dsm)$ for double-tag $\Dspm\to\Km\Kp\pipm$ candidates.  In the two-dimensional histogram, the narrow (blue) box around the mass peak illustrates the signal region, and the two wider (red) boxes further away from the peak illustrate the sideband region.}
\label{fig:Dsplots}
\end{center}
\end{figure}

CLEO determines single-tag yields by fitting $\MDs$ distributions for candidates that pass a very loose cut on $\Mbc$.  Double-tag yields are determined by counting the numbers of events in signal regions in the $M(\Dsm)$ vs. $M(\Dsp)$ plane, and subtracting backgrounds determined from the numbers of events in sideband regions.  These procedures are illustrated for $\Dspm\to\Km\Kp\pipm$ candidates in \Fig{fig:Dsplots}.

\begin{figure}[htb]
\begin{center}
\includegraphics[width=2.9in]{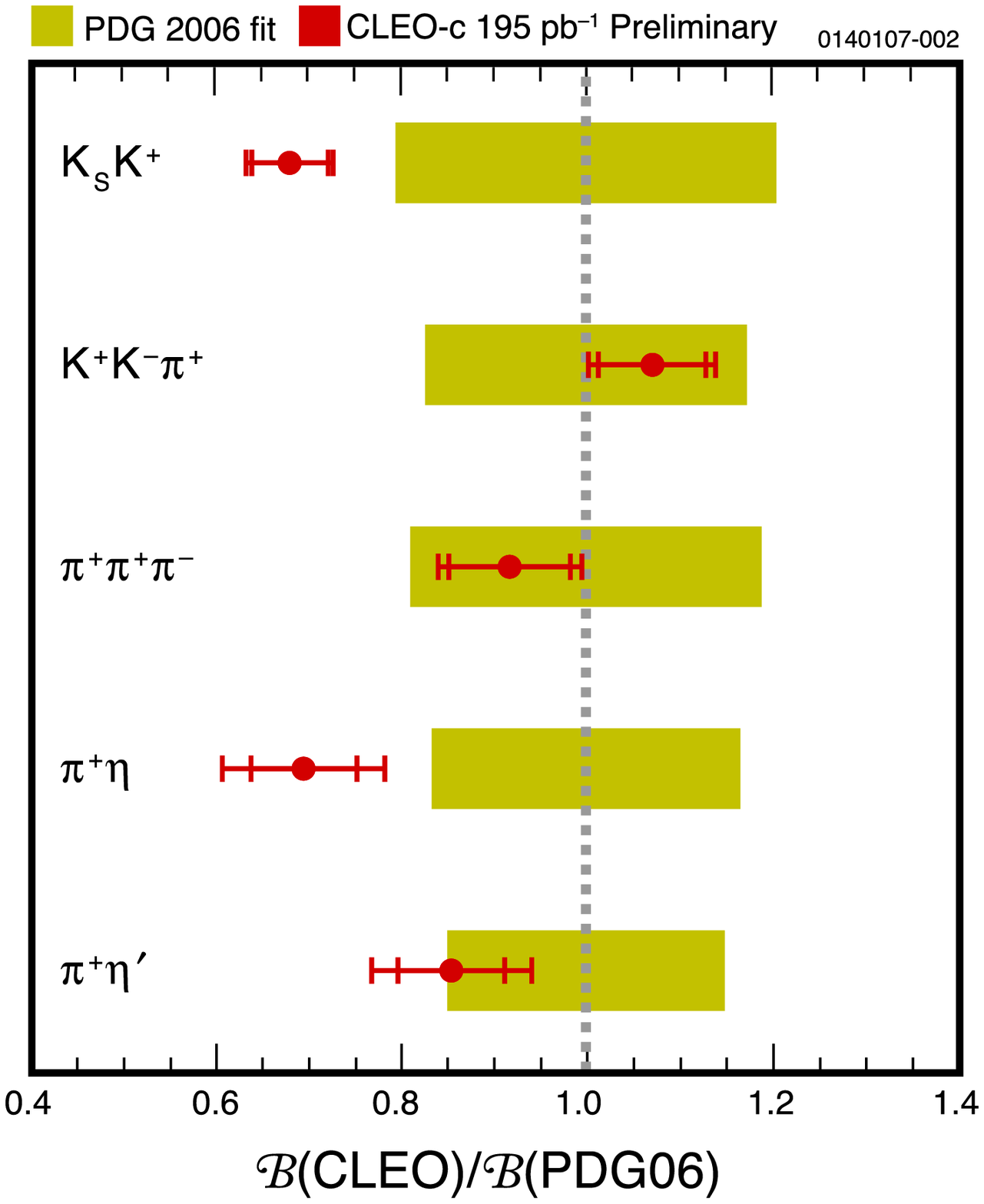}
\caption{The ratios $\calB(\textrm{CLEO})/\calB(\textrm{PDG06})$ of preliminary CLEO-c hadronic $\Dsp$ branching fractions to the PDG06 averages.  The widths of the PDG bars correspond to the errors in the averages.}
\label{fig:BDsp-PDG}
\end{center}
\end{figure}

The $\chi^2$ fit~\cite{wsun} procedure used in the measurements of absolute $\Dz$ and $\Dp$ branching fractions is applied to the ST and DT yields for six $\Ds$ decay modes.  Preliminary absolute branching fractions obtained from this analysis are given in \Tab{tab:BDs}, and these branching fractions are compared to the PDG06 averages in \Fig{fig:BDsp-PDG}.  The preliminary CLEO-c results are clearly significantly more precise than the PDG06 averages.  CLEO has an additional 130~\pbinv\ of data at this energy to be analyzed.  These data will be included in the publication of the 195~\pbinv\ data sample.
 
\begin{table}[htb]
\begin{center}
\Begtabular{l|c}
$\Dsp$ Mode & $\calB$ (\%) \\
\hline
$K_{S} K^+$ & $1.50 \pm 0.09 \pm 0.05$ \\
$K^- K^+ \pi^+$ & $5.57 \pm 0.30 \pm 0.19$\\
$K^- K^+ \pi^+ \pi^0$ ~~~~& $5.62 \pm 0.33 \pm 0.51$\\
$\pi^+ \pi^+ \pi^-$ & $1.12 \pm 0.08 \pm 0.05$ \\
$\pi^+ \eta$ & $1.47 \pm 0.12 \pm 0.14$ \\
$\pi^+ \eta'$ & $4.02 \pm 0.27 \pm 0.30$\\
\Endtabular
\caption{Preliminary absolute branching fractions for six hadronic $\Dspm$ decay modes measured by CLEO-c.}
\label{tab:BDs}
\end{center}
\end{table}

Belle measures $\calB(\Dsp\to\Km\Kp\pip)$ by applying a clever partial reconstruction technique to $\elp\elm \to\Dspmstar D_{s1}^\mp(2536)$ events~\cite{belle:dsabsolute} that produces double tag data samples.   A total of 552.3~\fbinv\ of data taken at the $\Upsilon(4S)$ were used in this analysis.  In the first tag, the $D_{s1}$ is fully reconstructed in the $D_{s1}\to\Dstar K$ mode, while only the $\gamma$ from $\Dsstar\to\Ds\gamma$ is observed.  In the second tag, the $\Dsstar$ is fully reconstructed in the decay chain $\Dspmstar\to\Dspm\gamma$ followed by $\Dspm\to\Km\Kp\pipm$, while only the kaon from $D_{s1}\to\Dstar K$ is observed.   Signal yields are determined by fitting missing mass distributions for the particles that are not reconstructed.  The yields from the first tag are proportional to the branching fractions for the $\Dstar$ decay chain, while the yields from the second tag are proportional to the branching fraction for $\Dspm\to\Km\Kp\pipm$.  Hence, $\calB(\Dsp\to\Km\Kp\pip)$ can be obtained from the ratio of the second tag yield to the first tag yield, using known branching fractions for the $\Dstar$ decay chain, and efficiencies.

The preliminary Belle and CLEO-c results are compared in \Tab{tab:belle-dsabsolute} to each other and to the PDG06~\cite{pdg:06} average.  With its large error, the PDG~06 value is consistent with the Belle and CLEO-c results.  However, the more precise Belle and CLEO-c are not very consistent with each other.  Hence, more work will be required to be confident that we know absolute $\Dsp$ branching fractions. 

\begin{table}[htb]
\begin{center}
\Begtabular{l|c}
  & $\calB(\Dsp\to\Km\Kp\pip)$ (\%) \\ \hline
Belle \hfill Preliminary  & $4.0~ \pm 0.4~ \pm 0.4~$ \\
CLEO-c  \hfill Preliminary & $5.57 \pm 0.30 \pm 0.19$ \\
PDG06                   & $5.2~~ \pm 0.9~$ \hspace*{2.5em} \\
\Endtabular
\caption{Comparison of preliminary Belle and CLEO-c measurements of the absolute branching fraction for $\Dsp\to\Km\Kp\pip$ decay with the PDG06 average.}
\label{tab:belle-dsabsolute}
\end{center}
\end{table}

\subsection{Partial $\Dsp \to \Km\Kp\pip$ Branching Fractions}

The branching fraction of the decay chain $\Dsp\to\phi\pip\to\Km\Kp\pip$ is one of the largest $\Dsp$ branching fractions, and -- indeed -- it was the decay sequence that provided the first observation of the $\Dsp$.  A branching fraction called $\calB(\Dsp\to\phi\pip)$ has often been used as the reference branching fraction for $\Ds$ decays.  As in the $\Dsp$ discovery, it is usually derived from a narrow mass cut around the $\phi$ peak in the $M(\Kp\Km)$ distribution in $\Dsp\to\Km\Kp\pip$ events.

\begin{figure}[htb]
\begin{center}
\includegraphics[width=3.3in]{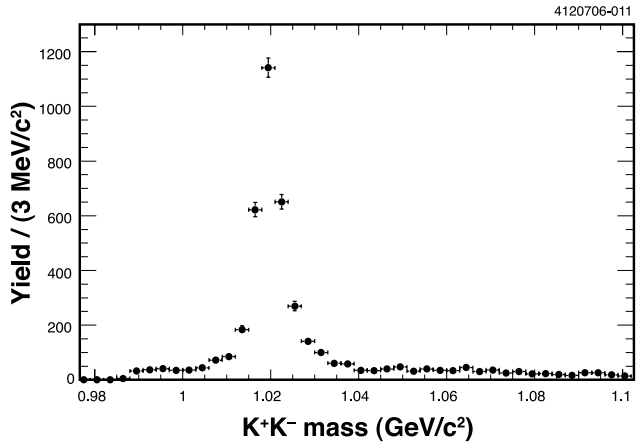}
\caption{The invariant mass distribution $M(\Kp\Km)$ for $\Dsp\to\Km\Kp\pip$ candidates in CLEO-c data.}
\label{fig:cleo-mkpkm}
\end{center}
\end{figure}

However, E687~\cite{E687:DsKKpi} has published and FOCUS~\cite{focus:DsKKpi} has reported significant contributions from $f_0(980)$ (or $a_0(980)$) in the $\phi\pi$ region of the $\Dsp\to\Km\Kp\pip$ Dalitz plot.  \Fig{fig:cleo-mkpkm} illustrates the invariant $\Kp\Km$ mass distribution for $\Dsp\to\Km\Kp\pip$ events.  A noticeable contribution (presumably scalar) is visible under the $\phi$ peak in in this distribution.  These contributions from other processes constitute approximately 5\% of the total yield in a reasonable mass range in $M(\Km\Kp)$ around the $\phi$ mass, $M_\phi$.  These contributions, which are not from $\Dsp\to\phi\pip$ decays, are comparable to current CLEO-c errors for partial branching fractions obtained using a narrow mass cut around $M_\phi$.  Hence, it is now difficult to make a case for identifying such a partial branching fraction with $\calB(\Dsp\to\phi\pip)$.  In response to this dilemma, CLEO now reports partial branching fractions, $\calB _{\Delta M} \equiv \calB(\Dsp\to\Km\Kp\pip)$ with $|M(\Km\Kp) - M_\phi| < \Delta M\textrm{ MeV}/c^2$.  Preliminary CLEO-c measurements with $\Delta M = 10$ and 20~MeV/$c^2$ are given in \Tab{tab:BdeltaM}.  I advocate using partial branching fractions like these for the reference branching fractions for other $\Dsp$ decays.  As $\Dsp$ branching fraction measurements become more precise, I hope that we can reach agreement on a reasonable choice for a $\Dsp$ reference branching fraction for most other decays that is not called $\calB(\Dsp\to\phi\pip)$, even though it has non-negligible contributions from other resonances in the $\Km\Kp\pip$ Dalitz plot.

\begin{table}[htb]
\begin{center}
\Begtabular{l|c}
~~~~~ & $\calB _{\Delta M}$ (\%) \\ \hline
$\calB_{10}$ & $1.98 \pm 0.12 \pm 0.09$ \\
$\calB_{20}$ & $2.25 \pm 0.13 \pm 0.12$ \\ \hline\hline
PDG06~~  & $1.77 \pm 0.44$\hspace*{3.0em} \\
\Endtabular
\caption{Preliminary partial branching fractions $\calB _{\Delta M} \equiv \calB(\Dsp\to\Km\Kp\pip)$ with $|M(\Km\Kp) - M_\phi| < \Delta M\textrm{ MeV}/c^2$ measured by CLEO. The PDG06 result is determined from $\calB(\phi\to\Km\Kp)$ and reported $\calB(\Dsp\to\phi\pip)$ values.\label{tab:BdeltaM}}
\end{center}
\end{table}

\section{\boldmath Inclusive $\Dz$, $\Dp$, and $\Ds$ Decays to $s\bar{s}$}

CLEO has measured branching fractions for inclusive $\Dz$, $\Dp$, and $\Ds$ decays to $\eta X$, $\eta' X$, and $\phi X$~\cite{cleo:Dssbar}.  Due to the $s\sbar$ content of $\eta$, $\eta'$, and $\phi$, we expect that the inclusive branching fraction for $\Ds$ decays to these particles would be larger than the branching fractions for corresponding $\Dz$ and $\Dp$ decays.  CLEO fully reconstructs one $D$ in a favorable mode used in the absolute measurements of hadronic branching fractions.  Selection criteria for these $D$ decay candidates include requirements on $\Delta E$ (see \Sec{sec:dabsolute}).  Then CLEO searches for $\eta$, $\eta'$ and $\phi$ in the decay products from the other $D$.  For this analysis, CLEO utilizes 281~\pbinv\ of $\psidprime$ data for $\Dz$ and $\Dp$ decays and 195~\pbinv\ of data at $E_{cm} \approx 4.17$~GeV for $\Ds$ decays.  

\begin{figure}[htb]
\begin{center}
\includegraphics[width=5in]{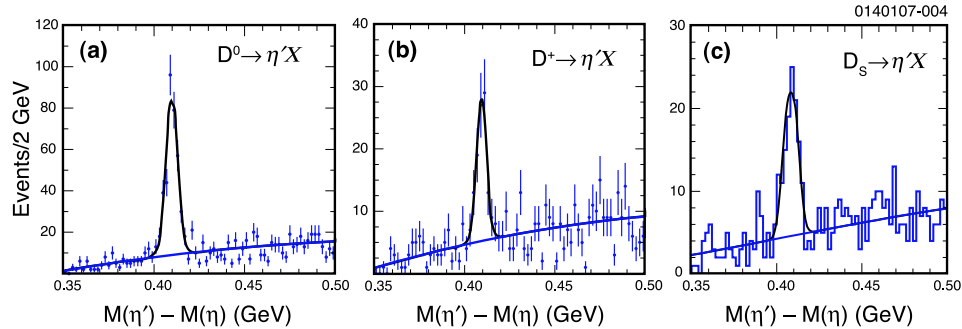}
\caption{Histograms of the mass differences between $\eta'$ and $\eta$ candidates with $\eta'\to\eta\pip\pim$ with $\eta\to\gamma\gamma$: (a) candidates for $\Dz\to\eta'X$ decay, (b) candidates for $\Dp\to\eta'X$ decay, and (c) candidates for $\Ds\to\eta'X$ decay.}
\label{fig:cleo-Detaprime}
\end{center}
\end{figure}

\Fig{fig:cleo-Detaprime} illustrates the data used to determine the inclusive $\eta' X$ signal yields.  For these inclusive modes, CLEO uses distributions in $M(\eta')-M(\eta)$, the difference between the invariant masses of the $\eta'$ candidate and the $\eta$ candidate in its decay chain.  Similar distributions for $M(\eta)$ and $M(\phi)$ are used to determine the signal yields for $\eta X$ and $\phi X$.  Fitted yields from $\Delta E$ sidebands are then subtracted from the signal yields.  The resulting inclusive branching fractions are given in \Tab{tab:Dssbar}.
These results lead to several qualitative observations including:
\Begitem
\item $\eta'$s and $\phi$s are relatively rare in $\Dz$ and $\Dp$ decay,
\item $\eta$s with their lower mass and larger light quark content are produced at substantially higher rates in $\Dz$ and $\Dp$ decays than $\eta'$ and $\phi$,
\item the $\phi$ rate in $\Ds$ decay is much higher than it is in $\Dz$ and $\Dp$ decay, and 
\item the higher $\phi$ rates in $\Ds$ decay can be used to separate $\Ds$ from $\Dz$ and $\Dp$ at $\Upsilon(5S)$ and hadron colliders.
\Enditem
CLEO has already used the observation in the last bullet to study the $B_s$ fraction in the final states in $\Upsilon(5S)$ decay~\cite{cleo:Y(5S):phi}.

\begin{table}[htb]
\begin{center}
\Begtabular{l|ccc}
\hline
Mode ~& $\calB(\Dz)$ (\%) & $\calB(\Dp)$ (\%) & $\calB(\Dsp)$ (\%) \\ \hline
$\eta  X$   &  $9.5  \pm 0.4 \pm 0.8$    & $6.3\pm 0.5 \pm 0.5$ & $23.5 \pm  3.1\pm 2.0$  \\
$\eta' X$  &  $2.48 \pm  0.17 \pm 0.21$ & $1.04 \pm 0.16 \pm 0.09$  & $8.7 \pm 1.9 \pm 0.8$  \\
$\phi  X$   &  $1.05 \pm 0.08 \pm 0.07$  & $1.03 \pm 0.10 \pm 0.07$  & $16.1 \pm 1.2\pm 1.1$  \\
\Endtabular
\caption{Branching fractions for inclusive $D\to s\bar{s} X$ decays measured by CLEO.\label{tab:Dssbar}}
\end{center}
\end{table}

\section{\boldmath Dalitz Analyses of Hadronic $D$ Decays}

In the last decade, large samples of hadronic $D$ decays have been available to the E791 and FOCUS collaborations from fixed target experiments at Fermilab, and to the CLEO, BaBar, and Belle collaborations from $\elp\elm$ collider experiments.  Dalitz analyses of these data samples have enabled new insights into the resonant substructure of multibody hadronic $D$ decays.   Historically, simply disentangling the resonant substructure has been the major emphasis of these studies.  More recently Dalitz analyses of $D$ decays have been stimulated by the possibility that the results can be used to determine the CKM matrix angles $\gamma$ or $\phithree$.

\subsection{\boldmath Dalitz Analyses of $\Dp\to\pip\pip\pim$ Decays}

It has long been understood that several $\pip\pim$ resonances are likely to contribute to the resonant substructure of $\Dp\to\pip\pip\pim$ decay.  These include $\rho^0(770)$, $f_0(980)$, $f_2(g)$, and possible other higher resonances.  On the other hand, there has been evidence (not always conclusive) that a lower mass S-wave $\pip\pim$ state called the $\sigma^0$ might exist and might contribute to $\Dp\to\pip\pip\pim$ decay.  Finding proof for or against the existence of this state and its contribution to $\Dp\to\pip\pip\pim$ decays has been a major concern of Dalitz analyses of this final state. 

The choice of the theoretical model used to fit the Dalitz plot has complicated interpretation of these analyses.  The isobar model, in which the decay amplitude is written as a sum of Breit-Wigner resonances for the various contributions, is perhaps the simplest to interpret~\cite{E791:Dalitz-Dppippippim,cleo:Dalitz-DzKmpippiz}.   However, there are technical difficulties with broad and overlapping resonances in this approach.  In principle the K-matrix formalism overcomes many of these difficulties~\cite{Pennington:2006}, and analyses using it have begun to appear~\cite{focus:Dalitz-Dppippippim}.  However, there is relatively little experience with the K-matrix formalism in Dalitz analyses of charm decay.

\subsubsection{\boldmath Dalitz Analysis of $\Dp\to\pip\pip\pim$ Decays from E791}

\begin{figure}[htb]
\begin{center}
\includegraphics[width=5.5in]{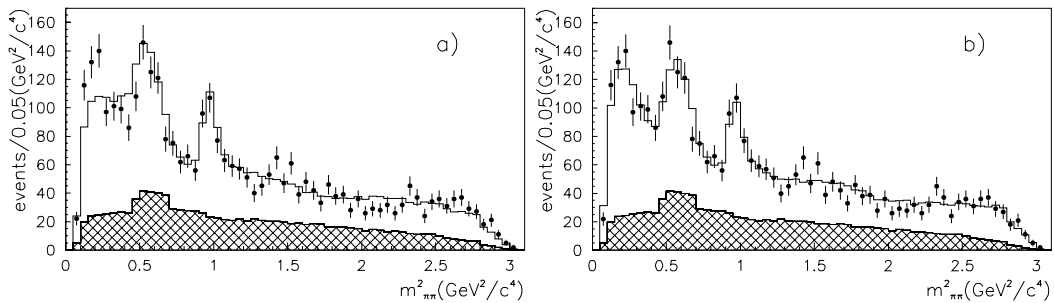}
\caption{Dalitz plot projections of the E791 data and fits for $\Dp\to\pip\pip\pim$ on the squares of invariant $\pi\pi$ masses: (a) for the  fit without a low mass scalar contribution and (b) for the fit with a low mass scalar contribution.  The histograms for the two $\pip\pim$ combinations are summed, and the shaded histograms are the background contributions.~\cite{aps:extra,aps:E791:Dalitz-Dppippippim}}
\label{fig:e791}
\end{center}
\end{figure}

The E791 Dalitz analysis of $\Dp\to\pip\pip\pim$ decays utilizes 1,686 candidates produced in 500~GeV/$c$\ $\pim$-nucleon interactions~\cite{E791:Dalitz-Dppippippim}.  In their isobar model they included contributions from the established resonances $\rho^0(770)$, $f_0(980)$, $f_2(1270)$, $f_0(1370)$, and $\rho^0(1450)$, and a non-resonant contribution.  The projection of the fit with these resonances onto the $m^2_{\pip\pim}$ axis is illustrated in \Fig{fig:e791}(a).  Clearly the fit does not represent well the data in the low mass region, $m^2_{\pip\pim} \approx 0.25$~GeV/$c^2$.  \Fig{fig:e791}(b) shows that the fit is much closer to the data when E791 includes a low mass scalar $\sigma^0$ contribution. The fit fractions obtained by E791 are given in \Tab{tab:Dalitz-E791&CLEO}.  It is interesting to note that the $\sigma^0$ contribution has the largest fit fraction.   

\begin{table}[htb]
\begin{center}
\Begtabular{l|cc}
 & ~~~CLEO (\%)~~~ & ~~~E791 (\%)~~~ \\ \hline
$\sigma^0\pip$    & $41.8 \pm 2.9$ & $46.3 \pm 9.2$ \\
$\rho^0\pip$      & $20.0 \pm 2.5$ & $33.6 \pm 3.9$ \\
$f_0(980)\pip$    & $4.1~ \pm 0.9$ & $6.2~ \pm 1.4$ \\
$f_2(1270)\pip$~~ & $18.2 \pm 2.7$ & $19.4 \pm 2.5$ \\
$f_0(1370)\pip$   & $2.6~ \pm 1.9$ & $2.3~ \pm 1.7$ \\
$f_0(1500)\pip$   & $3.4~ \pm 1.3$ & --- \\
Non Res           & $ < 3.5 $ & $7.8 \pm 6.6$ \\
$\rho(1450)\pip$  & $< 2.4 $ & $ 7.8~ \pm 0.6$ \\
\Endtabular
\caption{Fit fraction results of isobar Dalitz fits from E791 and CLEO for contributions to $\Dp\to\pip\pip\pim$ decays.  The E791 results were published, while the CLEO results are preliminary.}
\label{tab:Dalitz-E791&CLEO}
\end{center}
\end{table}

\subsubsection{\boldmath Dalitz Analysis of $\Dp\to\pip\pip\pim$ Decays from CLEO}
 
\begin{figure}[htb]
\begin{center}
\includegraphics[height=2.4in]{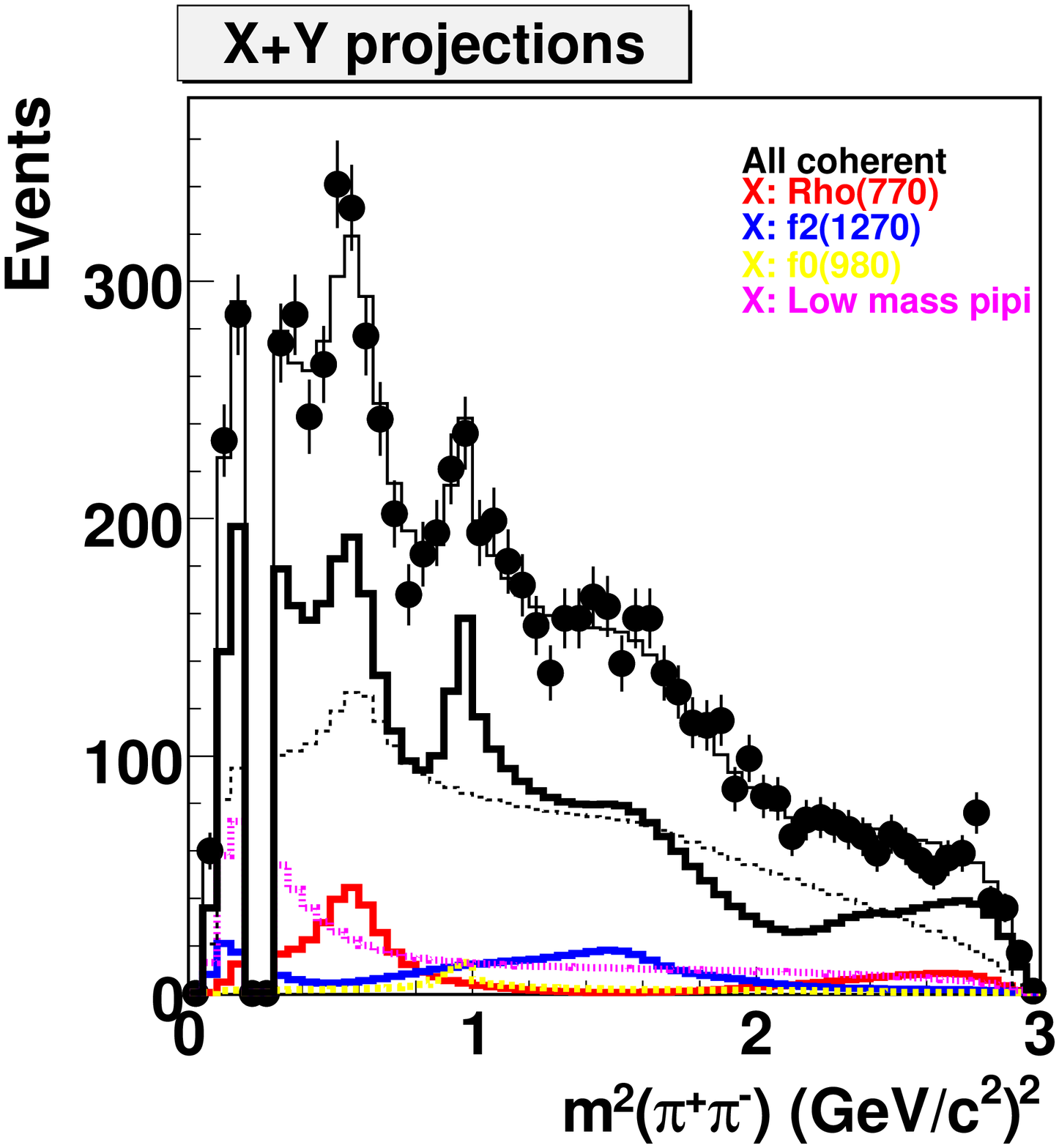}\hspace*{0.4in}
\includegraphics[height=2.4in]{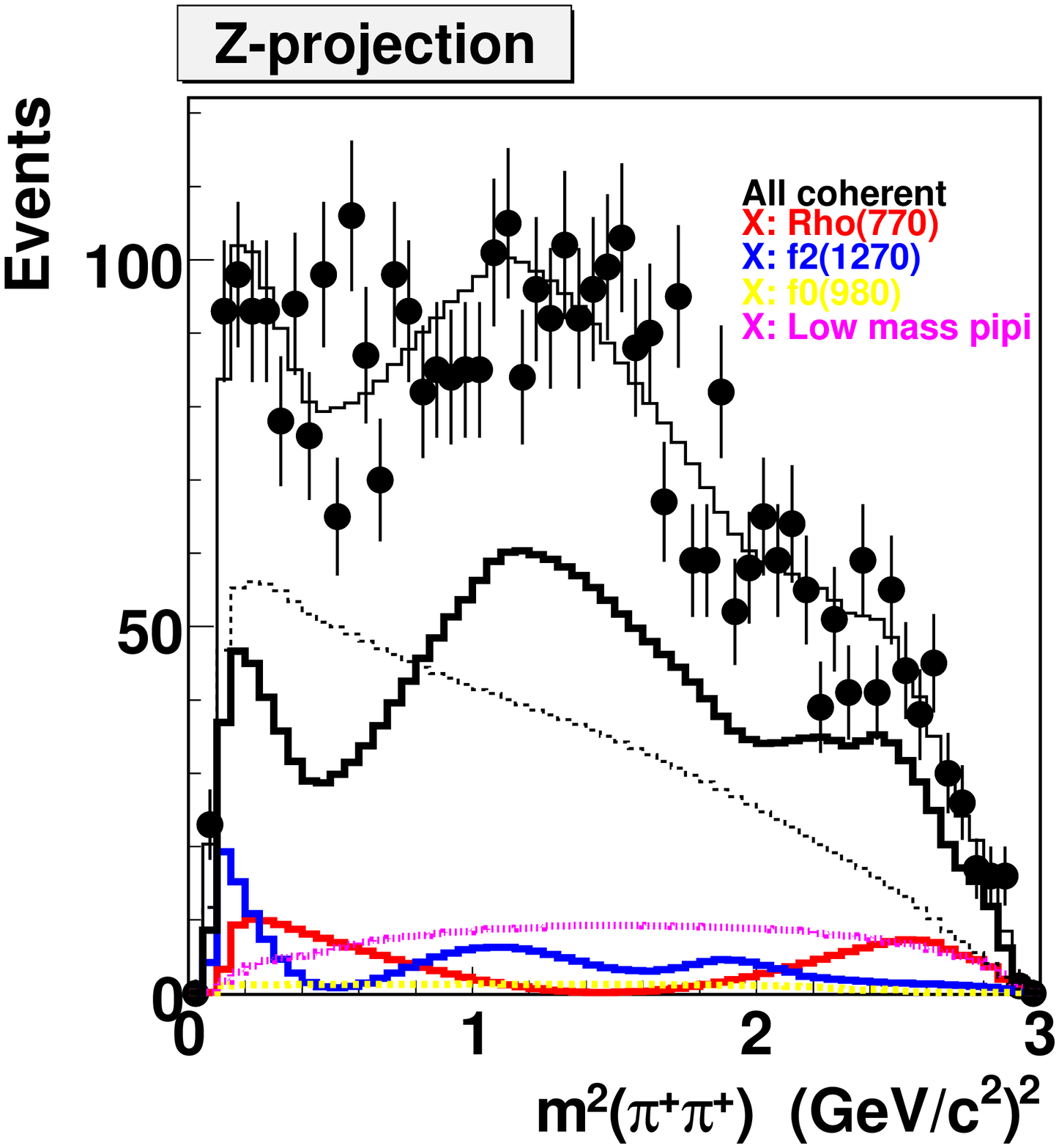}
\caption{Preliminary Dalitz plot projections on the squares of invariant $\pi\pi$ masses of the CLEO-c data and fits for $\Dp\to\pip\pip\pim$: (left) projections on the $\pip\pim$ invariant mass squared axis and (right) projections on the $\pip\pip$ invariant mass squared axis.  The empty bin at $m^2_{\pip\pim} = 0.25$~GeV/$c^2$ is due to a $\KS$ mass cut.}
\label{fig:cleo:Dalitz-Dppippippim}
\end{center}
\end{figure}

CLEO reported a preliminary Dalitz analysis of $\Dp\to\pip\pip\pim$ decay~\cite{cleo:Dalitz-Dppippippim} using an isobar model similar to the one used by E791. The CLEO-c data sample included 4,100 events with a signal to noise ratio of about two to one.   Projections on the isobar fit on the $m^2_{\pip\pim}$ and  $m^2_{\pip\pip}$ axes are illustrated in \Fig{fig:cleo:Dalitz-Dppippippim}.  The preliminary CLEO results are also listed in \Tab{tab:Dalitz-E791&CLEO}. Except for the $\rho(770)\pip$ contribution, the results of the two fits are in good agreement.   

\subsubsection{\boldmath Dalitz Analysis of $\Dp\to\pip\pip\pim$ Decays from FOCUS}
 
\begin{figure}[htb]
\begin{center}
\includegraphics[height=2.5in]{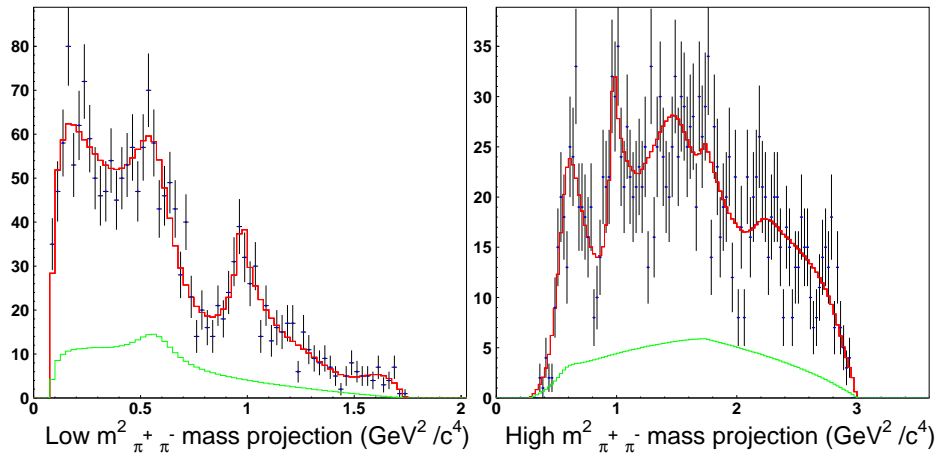}
\caption{Dalitz plot projections of the FOCUS data and fits for $\Dp\to\pip\pip\pim$ on the squares of invariant $\pi\pi$ masses: (a) projections on the low-mass $\pip\pim$ combination and (b) projections on the high-mass $\pip\pim$ combination.~\cite{plb:focus:Dalitz-Dppippippim}}
\label{fig:focus-Dalitz-Dppipipi}
\end{center}
\end{figure}

FOCUS reported~\cite{focus:Dalitz-Dppippippim} a Dalitz analysis performed on 1,527 $\Dp\to\pip\pip\pim$ decay candidates.  In the fit to the data, FOCUS included a nonresonant contribution, traditional isobar Breit-Wigner terms for 
$\rho^0(770)$ and $f_2(1270)$ contributions, and an S-wave contribution.  The S-wave contribution is not represented as a sum of scalar resonance contributions; it is represented by a K-matrix parameterization with five pole contributions.  The projections of the high and low mass combinations of the FOCUS data and fits on the $m^2_{\pip\pim}$ axis are illustrated in \Fig{fig:focus-Dalitz-Dppipipi}.  Perhaps the most illuminating comparisons of the FOCUS results with E791 and CLEO are the fit fractions of the $\rho^0(770)$ and $f_2(1270)$ contributions, which are $30.8 \pm 3.9$ and $11.7 \pm 1.9$, respectively.  Although there is qualitative agreement among the three experiments, a comprehensive understanding of the resonant substructure of $\Dp\to\pip\pip\pim$ will require substantially larger data samples and a better understanding of the strengths and weaknesses of the isobar and K-matrix formalisms.      
 
%\clearpage

\subsection{\boldmath Dalitz Analyses that Contribute to Measuring  $\gamma$ or $\phithree$}

Utilization of Dalitz analyses of $\Dz$ decays~\cite{babar:Dalitz-KSpippim,belle:Dalitz-KSpippim,cleo:Dalitz-KpKmpiz} to enable measurements of the CKM angle $\gamma$ or $\phithree$ is one of the more surprising and interesting applications of these analyses.  These techniques for measuring\footnote{To simplify notation, I will use $\gamma$ for either $\gamma$ or $\phithree$, or both.} $\gamma$ utilize $\Bpm\to\Dzhatstar\Kpm$ decays, where $\Dzhat$ can be $\Dz$ or $\Dzbar$ and the superscript $(*)$ denotes either $D$ or $\Dstar$.  The core idea~\cite{gamma:theory} is that the decay can be either Cabibbo favored (\eg, $\Bm\to\Dz\Km$) or doubly Cabibbo suppressed (\eg, $\Bm\to\Dzbar\Km$).  If the $\Dz$ and $\Dzbar$ decay to the same final state (\eg, $\Dzhat\to\KS\pip\pim$) the amplitudes for these two decays can interfere.  The phase of the interference term includes $\gamma$ and a strong phase; the former changes sign between $\Bm$ and $\Bp$ decays, while the latter does not.  Hence, this interference can generate a direct $CP$ violation, a difference between the rates of $\Bp\to\Dzhatstar\Kp$ and $\Bm\to\Dzhatstar\Km$ decays.  However, amplitude ratios and the strong phase in the $D$ decay are required to determine $\gamma$ from a measured $\Bpm$ decay rate asymmetry, and these parameters can be determined in a Dalitz analysis of $\Dz$ or $\Dzbar$ decays to the appropriate final state.   

\subsubsection{\boldmath Dalitz Analyses of $D\to\KS\pip\pim$ Decays to Measure $\gamma$}

\begin{figure}[htb]
\begin{center}
\includegraphics[height=1.9in]{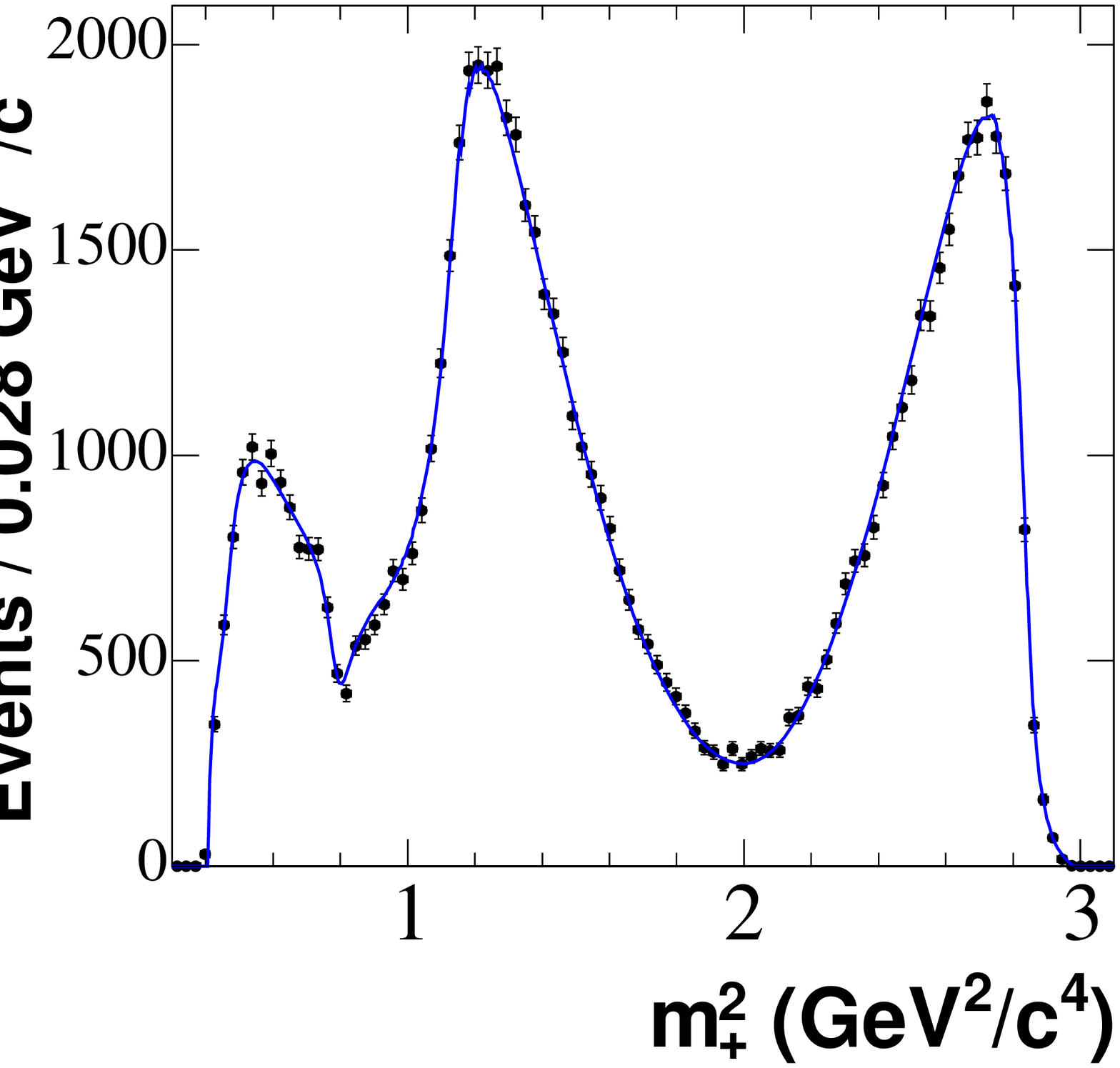}\hfill
\includegraphics[height=1.9in]{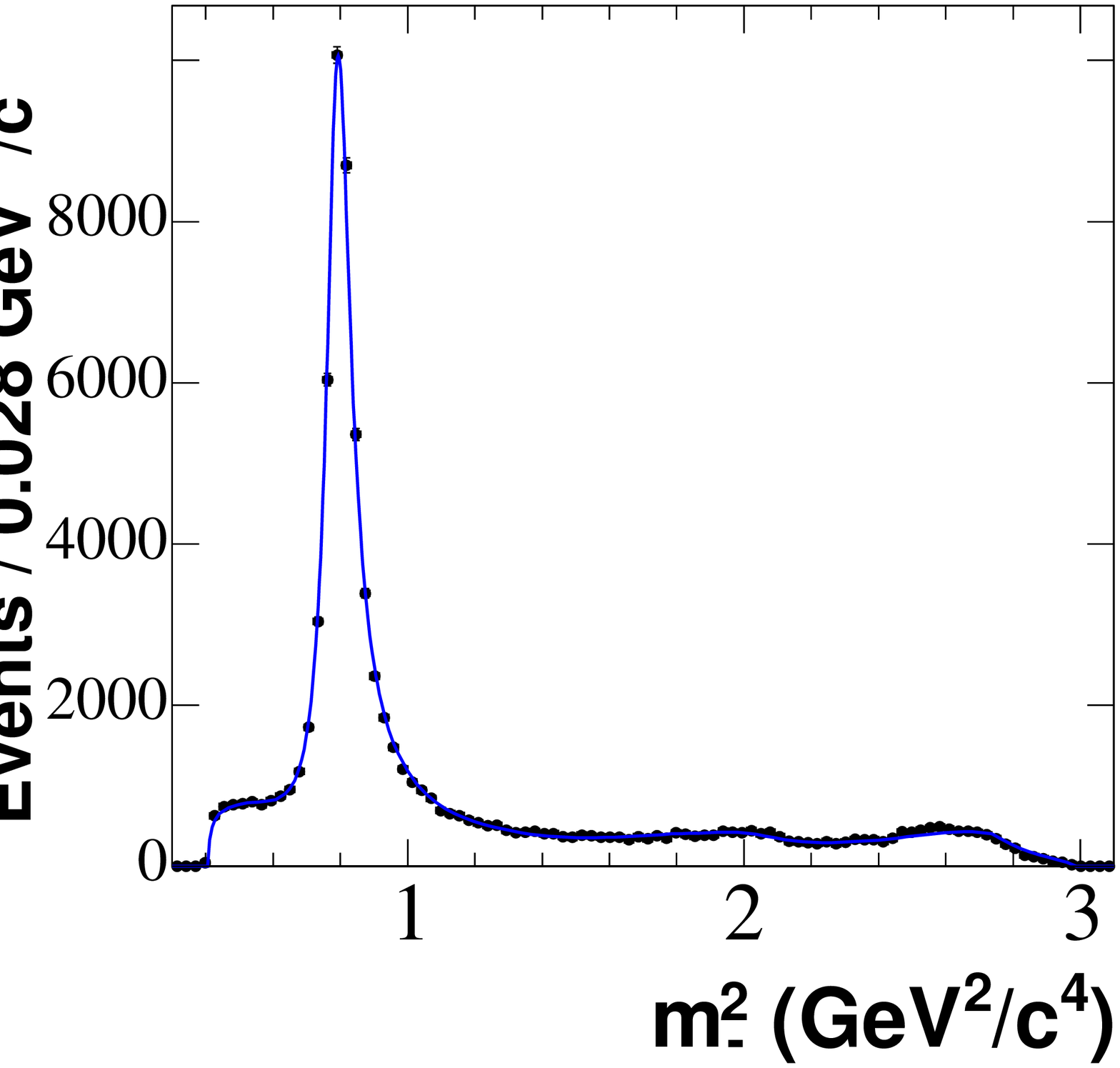}\hfill
\includegraphics[height=1.9in]{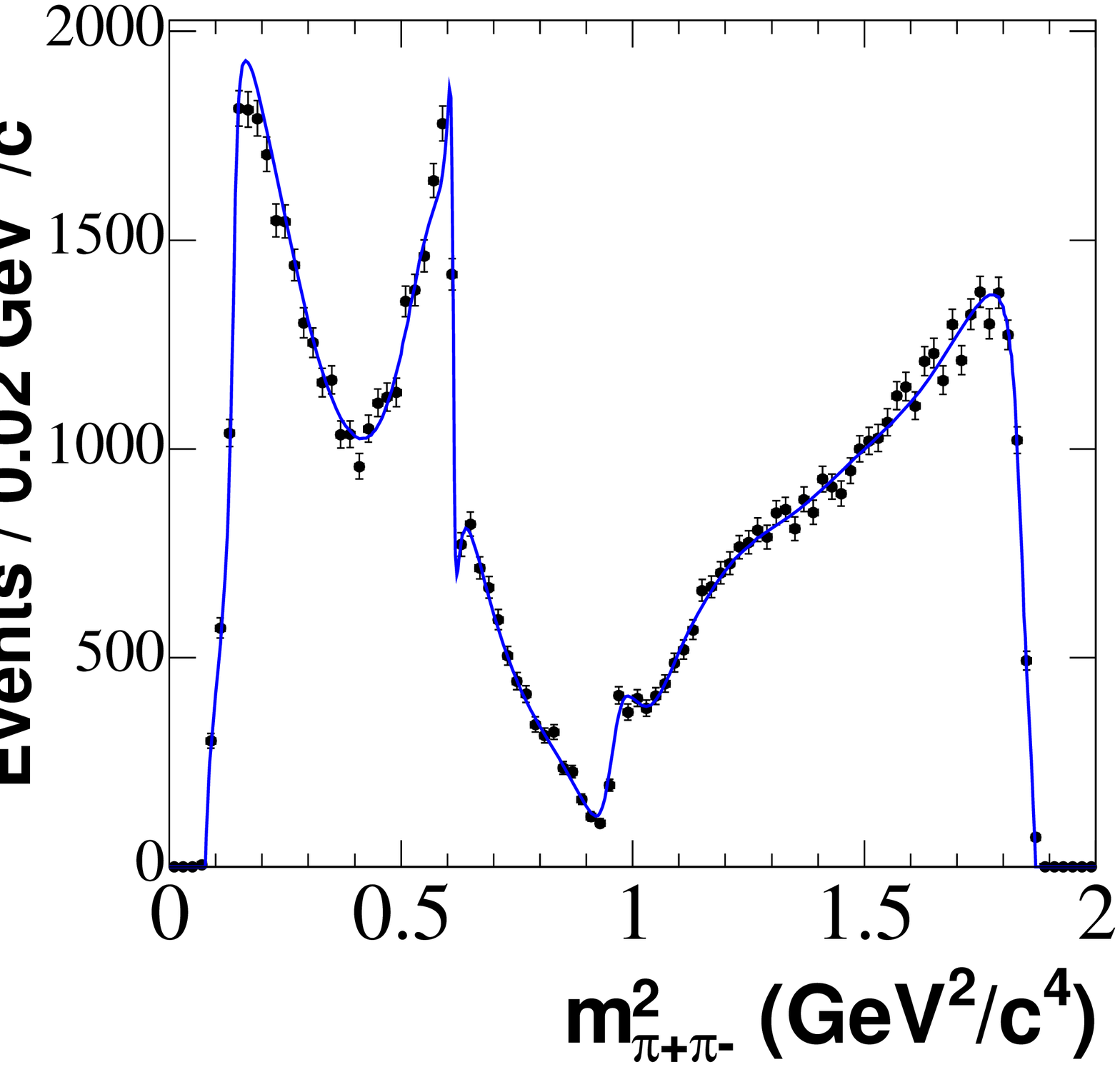}
\caption{Dalitz plot projections on the squares of invariant masses of the BaBar data and fits for $D\to\KS\pip\pim$: (left) projections on the $\KS\pip$ invariant mass squared axis, (middle) projections on the $\KS\pim$ invariant mass squared axis, and (right) projections on the $\pip\pim$ invariant mass squared axis.~\cite{aps:extra,aps:babar:Dalitz-KSpippim}}
\label{fig:babar:Dalitz-KSpippim}
\end{center}
\end{figure}

\begin{figure}[htb]
\begin{center}
\includegraphics[width=5.8in]{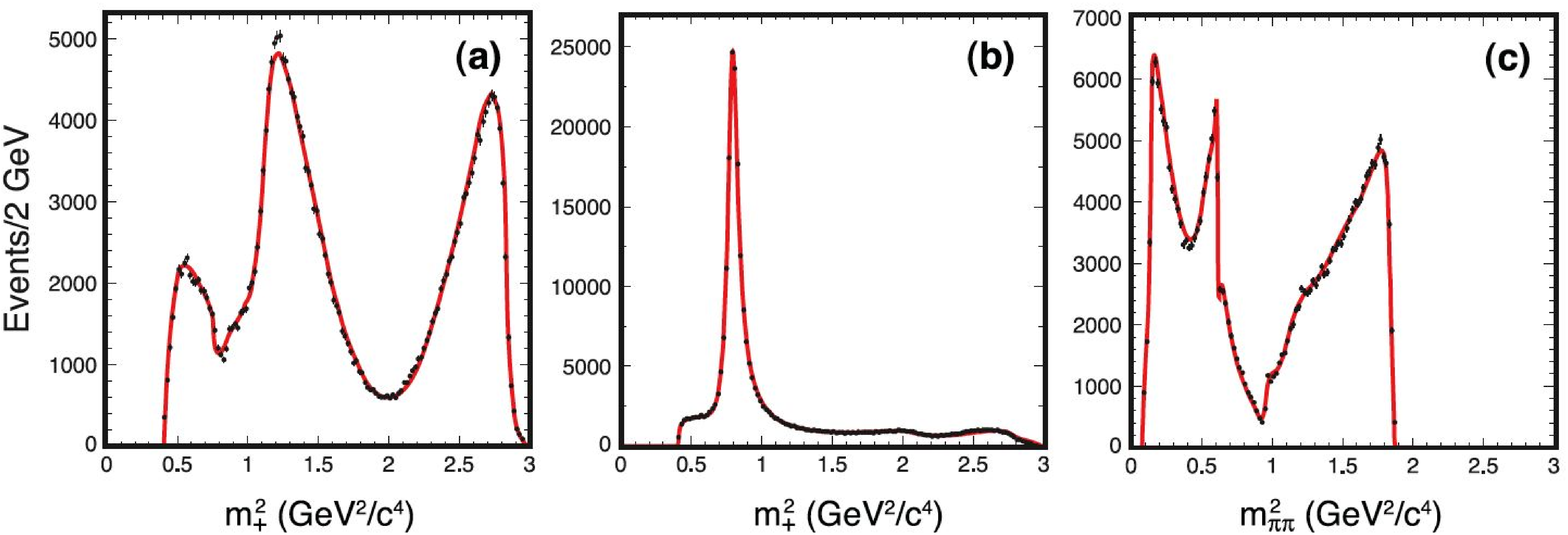}
\caption{Dalitz plot projections on the squares of invariant masses of the Belle data and fits for $D\to\KS\pip\pim$: (a) projections on the $\KS\pip$ invariant mass squared axis, (b) projections on the $\KS\pim$ invariant mass squared axis, and (c) projections on the $\pip\pim$ invariant mass squared axis.~\cite{aps:extra,aps:belle:Dalitz-KSpippim}}
\label{fig:belle:Dalitz-KSpippim}
\end{center}
\end{figure}

Recently, Babar~\cite{babar:Dalitz-KSpippim} and Belle~\cite{belle:Dalitz-KSpippim} analyzed the Dalitz plot for $\Dzhat\to\KS\pip\pim$ decays in $\Bpm\to\Dzhatstar\Kpm$ decays. For the Dalitz analyses, Belle utilized 262,000 signal events from a 357~\fbinv\ data sample, and Babar utilized 81,500 signal events from 91.5~\fbinv\ of data.  The purity of each final $\Dzhat$ data samples was approximately 97\%.  Both groups fit their Dalitz plots with isobar models; BaBar fit 16 two-body states and Belle fit 18 two-body states, and the BaBar states are a subset of the Belle states.  The projections of the data and fits on the Dalitz plot axes from the Belle and Babar are illustrated in Figures~\ref{fig:belle:Dalitz-KSpippim} and \ref{fig:babar:Dalitz-KSpippim}, respectively.
Due to the enormous data samples, the errors on the data points are very small.  
The ability of the fits to match the data so well is remarkable.  The fit fractions for the four most prominent resonant contributions are given in \Tab{tab:babar:belle:Dalitz-DKSpippim}.  Once again these analyses demonstrate that high quality charm decay measurements can be derived from the enormous BaBar and Belle data samples. 
 
From these analyses, Belle finds $\phithree = 53^\circ\ ^{+15^\circ}_{-18^\circ} \pm 3^\circ \pm 9^\circ$ and Babar finds $\gamma = 70^\circ \pm 31^\circ\ ^{+12^\circ}_{-10^\circ}\ ^{+14^\circ}_{-11^\circ}$, where the third errors are estimates of the uncertainties in the Dalitz decay models.  Within the large statistical errors there is good agreement between the two results.  Since the statistical errors are larger than the systematic errors, there is room for  improvement from the data that the two collaborations have not used in these analyses.  

\begin{table}[htb]
\begin{center}
\Begtabular{l|cc}
State & BaBar (\%) & Belle (\%) \\ \hline
$K^*(892)^+\pim$ & 58.6 & 61.2 \\
$\KS\rho^0$ & 22.4 & 21.6 \\
$\KS\sigma$ & 9.3 & 9.8 \\
Non Res & 7.3 & 9.7 \\
\Endtabular
\caption{The fit fractions for the four most prominent resonant contributions to $\Dzhat\to\KS\pip\pim$ obtained by BaBar and Belle.\label{tab:babar:belle:Dalitz-DKSpippim}}
\end{center}
\vspace*{-0.15in}
\end{table}

\subsubsection{\boldmath Dalitz Analysis of $\Dz\to\Kp\Km\piz$ Decays to Measure $\gamma$}

\begin{figure}[htb]
\begin{center}
\includegraphics[width=5.8in]{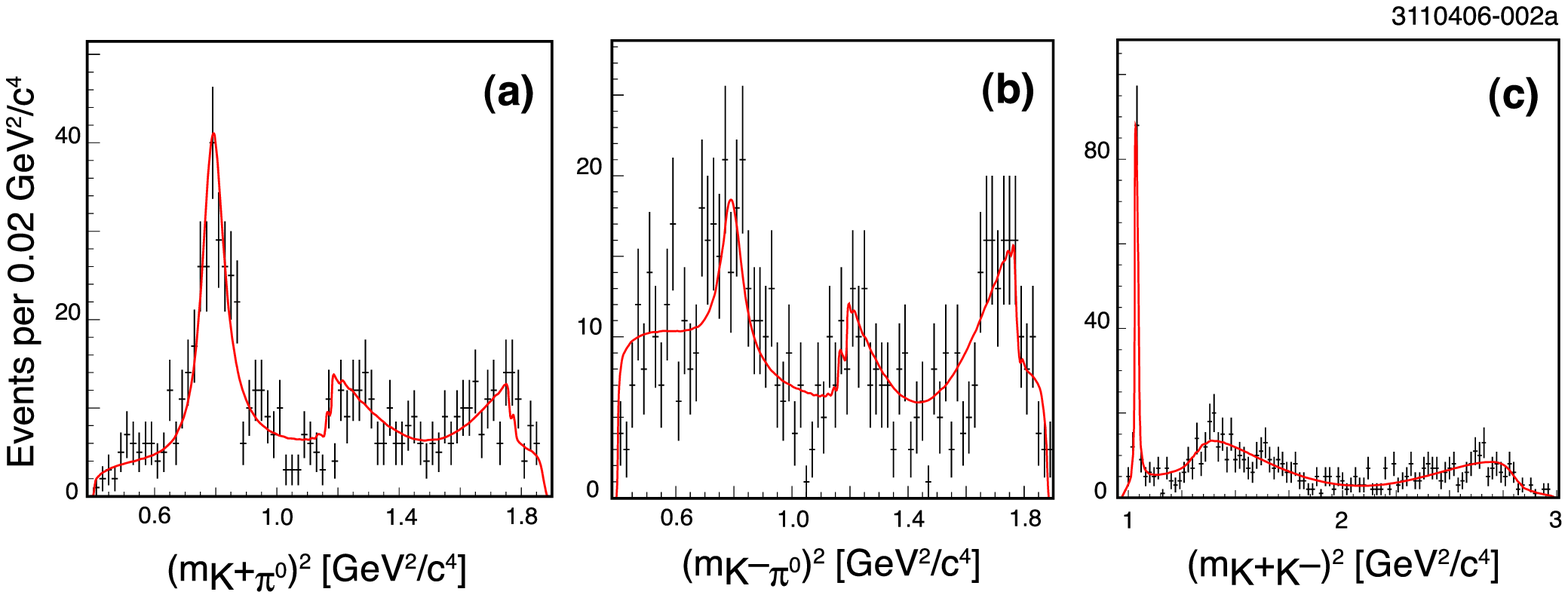}
\caption{Dalitz plot projections on the squares of invariant masses of the CLEO-c data and fits for $\Dz\to\Kp\Km\piz$: (a) projections on the $\Kp\pim$ invariant mass squared axis, (b) projections on the $\Km\piz$ invariant mass squared axis, and (c) projections on the $\Kp\Km$ invariant mass squared axis.}
\label{fig:cleo:Dalitz-KpKmpiz}
\end{center}
\end{figure}

CLEO has studied the Dalitz plot of $\Dz\to\Kp\Km\piz$ decays~\cite{cleo:Dalitz-KpKmpiz}, which can also be used to measure $\gamma$ in $\Bpm\to\Dzhat\Kpm$ decays.  The relative complex amplitude for $\Dz\to\Kpstar\Km$ and
$\Dz\to\Kpstar\Km$ decays is required to determine $\gamma$.  This relative amplitude is the same as that for the two decays $\Dz\to\Kmstar\Kp$ and $\Dzbar\to\Kpstar\Km$ (and their charge conjugates) assuming $CP$ conservation in these $\Dzhat$ decays.  CLEO found 735 $\Dz\to\Km\Kp\piz$ candidates in 9.0~\fbinv\ of data taken with the CLEO~III detector.  The Dalitz plot for the events were fit with 13 resonance components and a flat non resonant component.  Four of these components, $\Kpstar$, $\Kmstar$, $\phi$ and non-resonant had the largest fit fractions.  The projections of the fit on the three mass-squared axes are illustrated in \Fig{fig:cleo:Dalitz-KpKmpiz}.   The relative complex amplitude for the $\Kpmstar\Kmp$ decays is defined by,
\[ r_D e^{i\delta_D} \equiv {a_{\Kmstar\Kp} \over a_{\Kpstar\Km}}
e^{i(\delta_{\Kmstar\Kp} - \delta_{\Kpstar\Km})} \]
where the $a$'s are the real parts of the amplitudes and the $\delta$'s are the phase shifts.  The results for $r_D$ and $\delta_D$ are $r_D = 0.52 \pm 0.05 \pm 0.04$ and $\phi_D = 332^\circ \pm 8^\circ \pm 11^\circ$.

\vspace*{-0.2in}
\subsection{\boldmath Analysis of $\Dz\to\Kp\Km\pip\pim$ Decays from FOCUS}\label{sec:focus-Dalitz-KKpipi}

\begin{figure}[htb]
\begin{center}
\includegraphics[width=4.9in]{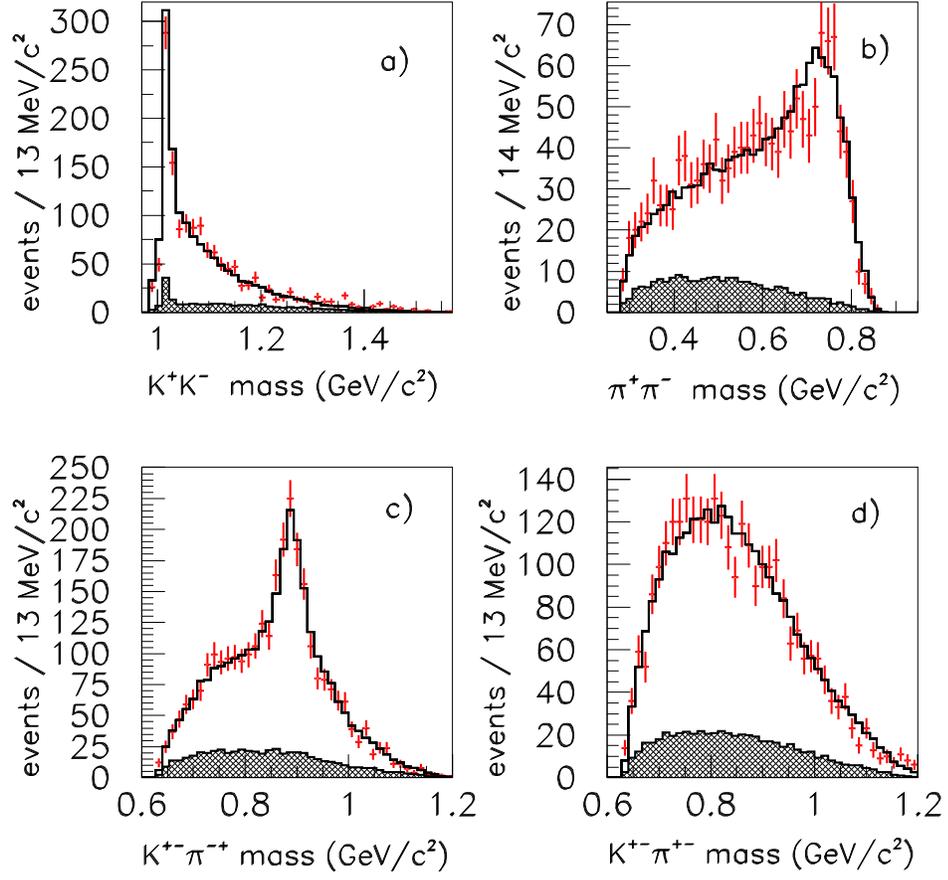}
\caption{Two-body invariant mass distributions from the FOCUS Collaboration's study the four-body decay $\Dz\to\Kp\Km\pip\pim$.~\cite{plb:focus:Dalitz-DzKpKmpippim}}
\label{fig:focus:Dalitz-DzKpKmpippim}
\end{center}
\end{figure}

A study of resonant substructure in $\Dz\to\Kp\Km\pip\pim$ decay~\cite{focus:Dalitz-DzKpKmpippim} from FOCUS is the only substantial effort to study resonances in four-body charm decays.  After all cuts, FOCUS obtained $1279 \pm 48$\ $\Dz\to\Kp\Km\pip\pim$ events above a modest background.  A total of 10 resonant/decay contributions were considered in the analysis; three of these were $K_1(1270)\Km$ where the $K_1$ decayed to states with other resonances, $\rho^0(770)\Kp$, $K^*_0(1430) \pip$, and $K^*(890) \pip$.  The results of the fits projected onto invariant mass axes are illustrated in \Fig{fig:focus:Dalitz-DzKpKmpippim}.  The fit fractions for the largest components are given in \Tab{tab:focus-DzKpKmpippim}.    

\begin{table}[htb]
\begin{center}
\Begtabular{l|c}
Mode & Fit Fraction (\%) \\ \hline
$K_1(1270)^+\Km$ & $33 \pm 6 \pm 4$ \\
$K_1(1400)^+\Km$ & $22 \pm 3 \pm 4$ \\
$\phi\rho^0$     & $29 \pm 2 \pm 1$ \\
$K^*(1400)^0\Kp\pim$ & $11 \pm 1 \pm 1$ \\
$f_0(980)\pip\pim$ & $15 \pm 3 \pm 2$ \\
\Endtabular
\caption{Fit fractions for the largest components in the FOCUS Dalitz analysis of $\Dz\to\Kp\Km\pip\pim$ decays.\label{tab:focus-DzKpKmpippim}}
\end{center}
\end{table}

\clearpage

\section{Summary and Conclusions}

The large data samples and high-quality detectors of the BaBar, Belle, and CLEO-c experiments have led to substantial advances in precision and discovery reach of charm physics.   The CLEO Collaboration, operating the CLEO-c detector in the charm threshold region, is measuring absolute $D$ hadronic branching fractions with unprecedented precision.  

CLEO reports preliminary results for $\Dz$ and $\Dp$ absolute branching fractions obtained from 281~\pbinv\ of $\elp\elm\to \psidprime\to\Dp\Dm$~or~$\Dz\Dzbar$ data.  These results are limited by systematic errors that are as low as $\ltsim 3$\%.  CLEO expects to reduce these systematic errors via an ongoing effort.  At this level of precision, final state radiation, whose effects are $\ltsim 2$\%, must be considered.  This is an interesting problem for the Particle Data Group, since most measurements of branching fractions do not take FSR into account.  

CLEO also reports  preliminary results for hadronic $\Ds$ branching fractions from 195~\pbinv\ of $\elp\elm$ annihilation data taken near $\Ecm = 4.17$~GeV, which is just above the  $\Dspm\Dsmpstar$ threshold.  These results, with errors generally below 10\%, are a substantial improvement over previous measurements, although they are limited by statistics.  CLEO has an additional 130~\pbinv\ of data at this energy to be analyzed, and plans to take more data at this energy in the future.  The resonant substructure of the $\Ds\to\Km\Kp\pip$ decay mode is becoming an issue, since one of the contributions to this mode is $\Ds\to\phi\pip$, which has often been used as a reference branching faction for other $\Ds$ decays.  There appears to be a significant contribution ($\sim 5$\%) from scalar resonances in a reasonable region in the $M(\Kp\Km)$ mass distribution around the $\phi$ peak.  Since the error of the CLEO measurements of $\calB(\Ds\to\Km\Kp\pip)$ are comparable to the scalar contribution under the $\phi$, we 
need to define a new reference branching fraction for $\Ds$ decays
 
In addition to the measurements of Cabibbo-favored branching fractions mentioned in the previous paragraphs, many new and accurate measurements of singly- and doubly-Cabibbo-suppressed branching fractions are appearing.  BaBar, Belle, and CLEO-c are active in these analyses.  By utilizing their enormous data samples,  BaBar and Belle are starting to dominate measurements of the ratios of branching fractions to reference-mode branching fractions.  
 
There is substantially increased activity in the Dalitz analyses of hadronic final states of $D$ decays.  However, there are serious ambiguities in these analyses due to uncertainties in how to specify the amplitude for the resonant contributions.  For example, E791, FOCUS, and CLEO have all reported Dalitz analyses of $\Dp\to\pip\pip\pim$.  E791 and CLEO utilize a standard isobar parameterization of the decay amplitude, while FOCUS uses a K-matrix description  which has some theoretical advantages. From these three experiments there is general agreement on the main resonant contributions to this decay, but there is important disagreement on the details, particularly on the presence of a $\sigma$ scalar contribution to the decay.  Developing a consensus on the proper treatment of resonant substructure with multiple overlapping states will become even more important as the larger data samples with higher purity become available.

BaBar and Belle report Dalitz analyses of $D\to\KS\pip\pim$ decays as byproducts of their studies of $\Bpm\to\Dzhatstar\Kmp$ decays to measure the CKM angle $\gamma$ or $\phithree$.  These collaborations are able to select enormous data samples with high purity from tagged $B$ decays.  CLEO reports a Dalitz analysis of $\Dz\to\Kp\Km\piz$ decays which can also be used in a determination of $\gamma$ or $\phithree$.  

FOCUS reported the first analysis of the resonant substructure of a four-body hadronic charm decay, $\Dz\to\Kp\Km\pip\pim$.  This result on a relatively modest data sample demonstrates that analyses of resonant substructure of charm decays to four or more bodies is a fertile field of research.

In the near future, we can look forward to many more exciting and precise results from the charm sector from the BaBar, Belle, and CLEO collaborations.  In the slightly farther future, we expect that the BESIII experiment operating in the charm threshold region at the BEPCII $\elp\elm$ storage ring in Beijing will open a new frontier in the precision and reach of charm decay experiments.

\section*{Acknowledgements}

I am delighted to acknowledge the contributions of my CLEO and CESR colleagues whose effort produced the CLEO results reported here and many other important measurements in the heavy flavor sector over the years.  This report is based upon work supported by the National Science Foundation under Cooperative Agreement No.~0202078.  It is a pleasure to acknowledge the generous hospitality of Prof.\ S. Paul and the staff of HQL06, and the cooperation of the editors of the proceedings.


\begin{thebibliography}{99}
\bibitem{aps:extra}
Readers may view, browse, and/or download material for temporary copying purposes only, provided these uses are for noncommercial personal purposes.  Except as provided by law, this material may not be further reproduced, distributed, transmitted, modified, adapted, performed, displayed, published, or sold in whole or in part, without prior written permission from the American Physical Society.
\end{thebibliography}
\end{document}